\documentclass[11pt]{article}
\usepackage[dvips]{graphics}
\input{amssym.def}
\input{amssym.tex}

\usepackage{xcolor}

\title{\Large\bf 
SCATTERING BY A PERFORATED SANDWICH PANEL:\\  METHOD OF RIEMANN SURFACES}
\author{Y.A. Antipov\\
Department of Mathematics, Louisiana State University,\\
Baton Rouge LA 70803, USA\\
}

\newcommand{\sgn}{\mathop{\rm sgn}\nolimits}

\newcommand{\I}{\mathop{\rm Im}\nolimits}
\newcommand{\R}{\mathop{\rm Re}\nolimits}
\newcommand{\const}{\mbox{const}}

\newcommand{\beqa}{\begin{eqnarray}}
\newcommand{\eeqa}[1]{\label{#1}\end{eqnarray}}
\newcommand{\bequ}{\begin{equation}}
\newcommand{\eequ}[1]{\label{#1}\end{equation}}

\newcommand{\Md}{\partial}

\newcommand{\Ga}{\alpha}
\newcommand{\Gb}{\beta}

\newcommand{\Ge}{\epsilon}

\newcommand{\Gf}{\phi}

\newcommand{\Gg}{\gamma}
\newcommand{\Gc}{\chi}

\newcommand{\Gk}{\kappa}

\newcommand{\Gl}{\lambda}
\newcommand{\Gn}{\eta}

\newcommand{\Gt}{\theta}

\newcommand{\Gr}{\rho}

\newcommand{\Gs}{\sigma}

\newcommand{\Go}{\omega}
\newcommand{\Gx}{\xi}
\newcommand{\Gy}{\psi}
\newcommand{\Gz}{\zeta}
\newcommand{\GD}{\Delta}
\newcommand{\GF}{\Phi}
\newcommand{\GG}{\Gamma}
\newcommand{\GL}{\Lambda}
\newcommand{\GP}{\Pi}

\newcommand{\GO}{\Omega}

\newcommand{\GY}{\Psi}

\newcommand{\CA}{{\cal A}}
\newcommand{\CB}{{\cal B}}
\newcommand{\CC}{{\cal C}}

\newcommand{\CF}{{\cal F}}

\newcommand{\CI}{{\cal I}}
\newcommand{\CJ}{{\cal J}}

\newcommand{\CL}{{\cal L}}

\newcommand{\CR}{{\cal R}}

\def\Ba{{\bf a}}
\def\Bb{{\bf b}}

\newcommand{\beq}{\begin{equation}}
\newcommand{\eeq}{\end{equation}}
\newcommand{\barr}{\begin{eqnarray}}
\newcommand{\earr}{\end{eqnarray}}
\newcommand{\beqn}{\begin{equation*}}
\newcommand{\eeqn}{\end{equation*}}
\newcommand{\barrn}{\begin{eqnarray*}}
\newcommand{\earrn}{\end{eqnarray*}}

\newcommand{\fr}{\frac}

\begin{document}
\maketitle


\begin{abstract}

The model problem of scattering of a sound wave by an infinite plane structure
 formed by a semi-infinite acoustically hard screen and a semi-infinite sandwich panel
perforated from one side and covered by a membrane from the other  is exactly solved. The model is governed 
by two Helmholtz equations for the velocity potentials in the upper and lower half-planes coupled
by the Leppington effective boundary condition and the equation of vibration of a membrane in a fluid.
Two methods of solution are proposed and discussed. Both methods reduce the problem to an order-2
vector Riemann-Hilbert problem. The matrix coefficients have different entries, have the Chebotarev-Khrapkov structure
and share the same order-4 characteristic
polynomial.  Exact Wiener-Hopf matrix factorization requires solving  a scalar Riemann-Hilbert  on an elliptic surface
and the associated genus-1 Jacobi inversion problem solved  in terms of the associated Riemann $\Gt$-function. Numerical results for the absolute value
of the total velocity potentials are reported and discussed.

\end{abstract}

\setcounter{equation}{0}

\section{Introduction}

The effect of perforation on the transmission of sound waves through single-leaf and double-leaf  panels
was analyzed in   {\bf(\ref{ffo})}, {\bf(\ref{lep0})}. When  an elastic double-leaf honeycomb panel is perforated from one
or both sides
the transmission of sound is significantly reduced {\bf(\ref{lep})}. Leppington   {\bf(\ref{lep})} applied the method of matched asymptotic expansions
to analyze this effect for an infinite honeycomb cellular structure in the cases of acoustically hard or  acoustically
transparent cell walls. One of the main results of this work is the derivation of the effective boundary conditions.
In particular, in the case of cells with acoustically hard walls, the effective boundary condition on the panel surface $S=\{x\in I, 
y=0\}$
has the form
\beq
\Gy_{1y}-\Gy_{0y}+k\tau\Gy_1=0, \quad (x,y)\in S,
\label{1.1}
\eeq
where $\Gy_0$ and $\Gy_1$ are the velocity potentials in the lower and  upper half-planes, $\Gy_{jy}$ is the normal
derivative of $\Gy_j$, $k$ is the wave number,  and $\tau$ is a parameter that accounts for perforations.
The condition (\ref{1.1}) is to be complemented by the classical boundary condition
 \beq
 D_x[\Gy_{0y}]-\Ga(\Gy_1-\Gy_0)=0, \quad (x,y)\in S,
\label{1.2}
\eeq
where $D_x$ is the differential operator with respect to $x$ of order 2 or 4 depending whether the lower unperforated 
skin of the elastic structure is a membrane or an elastic plate and $\Ga$ is a parameter.
 
The model problem of scattering of a plane sound wave by an infinite plane structure formed by a semi-infinite acoustically hard screen and
semi-infinite honeycomb elastic panel with acoustically hard walls perforated from one side was reduced {\bf(\ref{jon})} to a Riemann-Hilbert
problem for two pairs of functions. This work does not factorize the matrix coefficient. Instead, it applies 
the asymptotic method of  small $\tau$ with the leading-order term determined by the decoupled problem.

The vector Riemann-Hilbert  {\bf(\ref{jon})}  was analyzed  {\bf(\ref{as1})} by the method of factorization
on a Riemann surface {\bf(\ref{moi})}, {\bf(\ref{am})}.  It was shown that the matrix coefficient of the vector Riemann-Hilbert
problem has the Chebotarev-Khrapkov structure {\bf(\ref{che})}, {\bf(\ref{khr})} with the characteristic polynomial $f(s)$
of degree 8. 
The Khrapkov methodology is applicable when $deg f(s)\le 2$. 
For a particular case of the Chebotarev-Khrapkov matrix and when $\deg f(s)=4$, a method of elliptic
functions eliminating the essential singularity at infinity was proposed in {\bf(\ref{dan})}. A numerical
approach of Pad\'e approximants  
for factorization of the Chebotarev–Khrapkov matrix was developed in {\bf(\ref{abr})}.
If $\deg f(s)=8$, then the exact representation of the  Wiener-Hopf matrix factors includes exponents of functions having 
an order-3 pole at infinity and therefore 
has an  unacceptable essential singularity. This singularity was eliminated {\bf(\ref{as1})}  by reducing the matrix factorization problem to a scalar Riemann-Hilbert problem on a hyperelliptic surface {\bf(\ref{zve})} and solving the associated
genus-3 Jacobi inversion problem   {\bf(\ref{kra})},  {\bf(\ref{spr})}. The solution  {\bf(\ref{as1})} was designed for the case of real wave numbers. In this case 
all eight branch points lie on the same circle and are symmetric with respect to the origin. That is why the brunch cuts and $A$- and $B$ - cross-sections are two-sided arcs of a circle. The full solution of the model problem
requires to determine two unknown constants by satisfying the additional conditions, to analyze the behavior of the Wiener-Hopf matrix factors at infinity, and based on the solution obtained develop an efficient numerical procedure for the velocity potentiails.
These aspects were not a scope of the investigation {\bf(\ref{as1})}.  

In this work we analyze the model problem of scattering of a plane sound wave by a structure similar to the one considered
in {\bf(\ref{jon})}, {\bf(\ref{as1})}. The only one difference is that the lower skin of the structure is  a membrane,  not  a thin plate. Mathematically, this means that instead of the fourth order differential operator $D_x$ in (\ref{1.2})
we have a second order operator. We aim to derive the full solution, find unknown constants, analyze the behavior of the matrix factors
at infinity and obtain numerical results.
In Section 2, we formulate the model problem and write down the governing boundary value problem for two Helmholts 
equations coupled by the boundary conditions. We apply the Laplace transform and convert the problem into an order-2  vector
Riemann-Hilbert problem in Section 3. We show that the problem of matrix factorization is equivalent to a scalar Riemann-Hilbert
problem on an elliptic surface.  To eliminate the essential singularity 
of the factors at infinity, we solve a genus-1 Jacobi inversion problem in terms of the Riemann $\Gt$-function. 
At the end of Section 3, we compute the partial indices of factorization and show that they are stable.
In Section 4, we propose another method of reduction of the model problem to a vector Riemann-Hilbert problem in order
to understand if the method has an advantage over the method applied in Section 3. Section 5 presents numerical results
for the velocity potentials  obtained on the basis of the solution derived in Section 3.

\setcounter{equation}{0}

\section{Formulation}\label{form}

Suppose that a semi-infinite sandwich panel is attached to a semi-infinite acoustically rigid 
screen (Fig. 1). The upper side of the panel  $M_1=\{0<x<\infty, y=d\}$  is perforated, while the lower side 
 $M_0=\{0<x<\infty, y=0\}$ is a smooth membrane. The unperforated and perforated skins are linked by cells whose
 sides are assumed to be acoustically hard. The rigid screen $S=\{-\infty<x<0, y=0\}$ and the lower side of the
 sandwich panel are clamped, and without loss of generality, the displacement is equal to zero at 
the junction point $x=0$, $y=0$.

\begin{figure}[t]
\centerline{
\scalebox{0.5}{\includegraphics{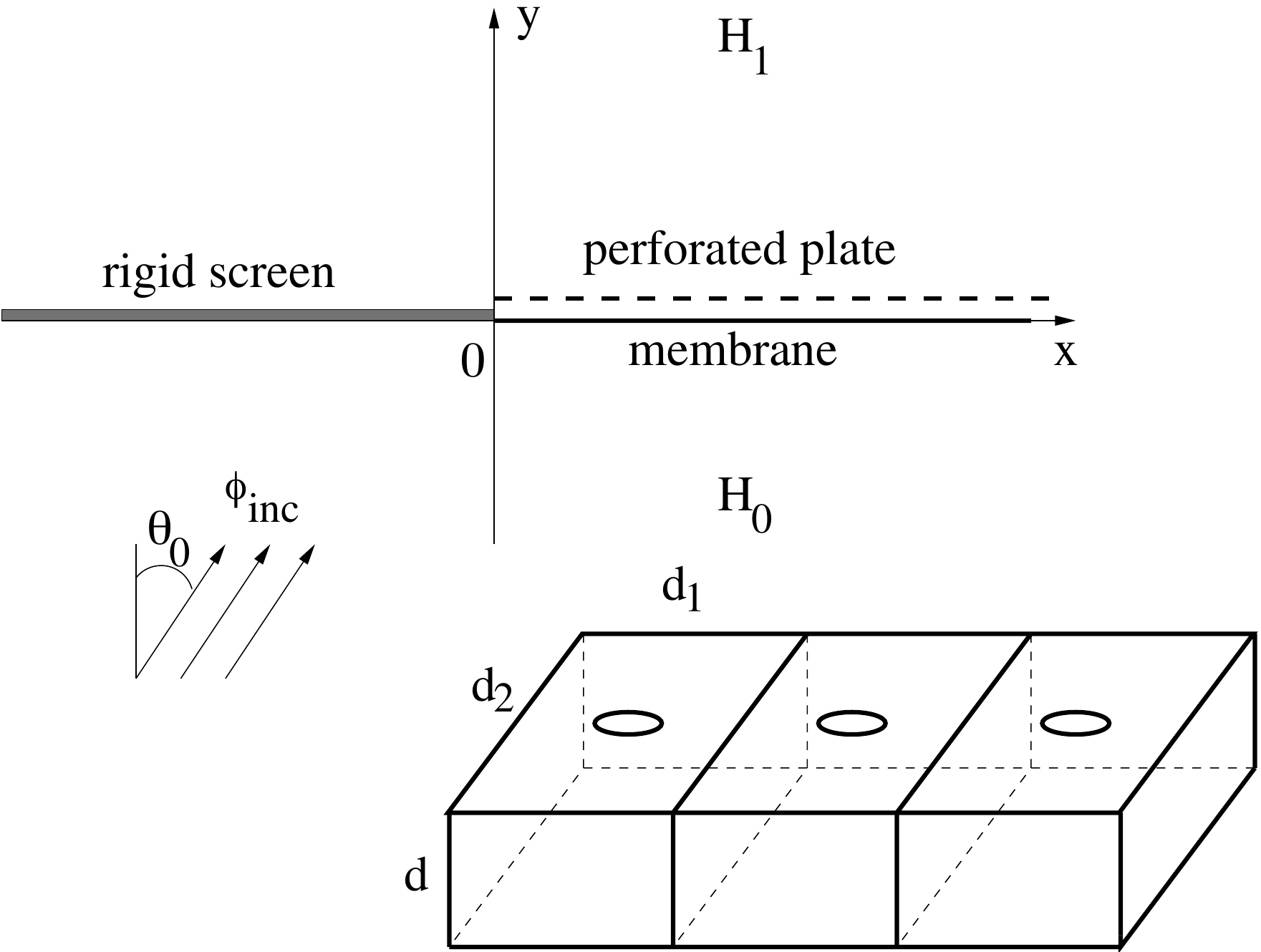}}
}
\caption{Sandwich panel attached to an acoustically rigid screen}
\label{fig1}
\end{figure}

Compressible fluid of wave speed $c$ occupies the regions outside the sandwich panel  and
the screen. The system is excited by a plane wave of incident velocity potential
\beq
\R  \{\Gf_{inc} e^{-i\Go t}\}=\R\{e^{ik(x\sin\Gt_0+y\cos\Gt_0}e^{-i\Go t}\}, \quad y<0,
\label{2.1}
\eeq
where $\Gt_0\in(-\pi/2,\pi/2)$, $k=\fr{\Go}{c}$ is the acoustic wave number, $k=k_1+ik_2$, $0<k_2<<k_1$, and $\Go$ is the radian frequency.
The velocity potentials  $\Gy_{0,1}e^{-i\Go t}$ in the lower and upper half-planes ${\Bbb H_0}$ and ${\Bbb H_1}$ satisfy the Helmholtz 
equation
\beq
\left(\fr{\Md^2}{\Md x^2}+ \fr{\Md^2}{\Md x^2}+k^2\right)\Gy_j(x,y)=0, \quad (x,y)\in{\Bbb H_j}, \quad j=0,1.
\label{2.2}
\eeq
The two potentials satisfy the standard acoustically hard boundary conditions
\beq
\fr{\Md \Gy_0}{\Md y}=\fr{\Md \Gy_1}{\Md y}=0, \quad x<0, \quad y=0.
\label{2.3}
\eeq
The surface deflection $\Gn_0 e^{-i\Go t}$ of the membrane $M_0$  and the pressure fluctuation $p_j e^{-i\Go t}$
are given in terms of the potentials by the relations
\beq
\Gn_0=\fr{i}{\Go}\fr{\Md\Gy_0}{\Md y}, \quad p_j=i\Go\Gr_f\Gy_j, \quad j=0,1,
\label{2.3'}
\eeq
where $\Gr_f$ is the mean density of the fluid.
The deflection of the membrane 
responds to the total surface
pressure
according to the linearized equation {\bf(\ref{pap})}
\beq
m_0\fr{\Md^2}{\Md t^2}\left(\Gn_0  e^{-i\Go t}\right)=T\fr{\Md^2}{\Md x^2} \left(\Gn_0 e^{-i\Go t}\right)-(p_1-p_0) e^{-i\Go t}, \quad y=0, \quad x>0.
\label{2.4}
\eeq
Here, $m_0$ is the mass per unit area of the membrane,  $T$ is the tension per unit length. Equivalently, this boundary condition  can be written in the form
\beq
\fr{\Md^3\Gy_0}{\Md x^2\Md y}+\mu^2\fr{\Md \Gy_0}{\Md y}-\Ga(\Gy_1-\Gy_0)=0, \quad y=0, \quad x>0,
\label{2.5}
\eeq
where 
\beq
\mu=\sqrt{\fr{m_0}{T}}\Go, \quad \Ga=\fr{\Gr_f\Go^2}{T}.
\label{2.6}
\eeq
The second boundary condition of the sandwich panel is the Leppington {\bf(\ref{lep})}
effective boundary condition
\beq
\fr{\Md \Gy_1}{\Md y}-\fr{\Md \Gy_0}{\Md y}+k\tau\Gy_1=0, \quad y=0, \quad x>0,
\label{2.7}
\eeq
where
\beq
\tau=\fr{kd}{1-k^2V/(2a)},
\label{2.8}
\eeq
where $a$ is the aperture radius and $V$ is the cell volume, $V=d_1d_2d$.  In the derivation {\bf(\ref{lep})} of the boundary
condition (\ref{2.7}), it is assumed that the wavelength is large
compared with the spacing parameters $d_1$, $d_2$ and $d$ (Fig. 1), so that 
$|k|d_1<<1$, $|k|d_2<<1$, and $|k|d<<1$. Since $|k|d$ is small, the boundary condition can be applied at $y=0$ to leading order. 
The parameter $\tau$ is dimensionless, while
the parameters $\mu$ and $\Ga$ have dimensions $L^{-1}$ and $L^{-3}$, respectively, with $L$ measured in units of length.
The condition $\tau=\infty$, an analog of the Helmholtz resonance  {\bf(\ref{dow})} condition, gives the critical value $k_{res}$  {\bf(\ref{lep})} 
of the wave number
\beq
k_{res}=\sqrt{\fr{2a}{V}}.
\label{2.8'}
\eeq
For complex wave numbers $k$, this condition is never satisfied. However, it is clear that
when $\R k>> \I k>0$ and $|k|\to k_{res}$, $|\tau|\to \infty$.

It is convenient to introduce a reflected wave of potential
$\Gf_{ref}=e^{ik(x\sin\Gt_0-y\cos\Gt_0)}$
and scattering potentials $\Gf_0$ and $\Gf_1$. Then the total velocity potentials are expressed through the incident, reflected
and scattering potentials by
\beq
\Gy_0=\Gf_{inc}+\Gf_{ref}+\Gf_0, \quad y<0; \quad \Gy_1=\Gf_1, \quad y>0.
\label{2.9}
\eeq 
The scattering potentials satisfy the Helmholtz equation
\beq
\left(\fr{\Md^2}{\Md x^2}+ \fr{\Md^2}{\Md x^2}+k^2\right)\Gf_j(x,y)=0, \quad (x,y)\in {\Bbb H_j},\quad j=0,1,
\label{2.10}
\eeq
and the  following boundary conditions:
$$
\fr{\Md \Gf_0}{\Md y}=\fr{\Md \Gf_1}{\Md y}=0, \quad x<0, \quad y=0,
$$$$
\fr{\Md^3\Gf_0}{\Md x^2\Md y}+\mu^2\fr{\Md \Gf_0}{\Md y}-\Ga(\Gf_1-\Gf_0)=-2\Ga e^{ikx\sin\Gt_0},  \quad x>0, \quad y=0,
$$
\beq
\fr{\Md \Gf_1}{\Md y}-\fr{\Md \Gf_0}{\Md y}+k\tau\Gf_1=0,  \quad x>0, \quad y=0.
\label{2.11}
\eeq
It is also required that the scattering potentials $\Gf_0$ and $\Gf_1$ satisfy an outgoing wave radiation condition as $x^2+y^2\to\infty$.

\setcounter{equation}{0}
  
\section{Vector Riemann-Hilbert problem}\label{vec}  

\subsection{Derivation based on the Laplace transform of the velocity potentials}

To reduce the boundary value problem (\ref{2.10}), (\ref{2.11}) to a vector Riemann-Hilbert problem
(known also as a matrix Wiener-Hopf problem), we introduce the Laplace transforms of the velocity potentials
\beq
\GF_{j+}(s;y)=\int_0^\infty \Gf_j(x,y)e^{i s x}dx, \quad \GF_{j-}(s;y)=\int_{-\infty}^0 \Gf_j(x,y)e^{i s x}dx.
\label{3.1}
\eeq
Their sums $\GF_{j}(s;y)=\GF_{j+}(s;y)+\GF_{j-}(s;y)$ satisfy the equations
\beq
\left[\fr{d^2}{dy^2}-(s^2-k^2)\right]\GF_{j}(s;y)=0, \quad j=0,1.
\label{3.2}
\eeq
Fix a single branch of the function $\Gg^2(s)=(s^2-k^2)$  in the $s$-plane
cut along the line joining the branch points $k$ and $-k$ and passing through the infinite point.
Denote by $\Gt_*=\arg k\in (0,\fr{\pi}{2})$ and select
\beq
-\pi+\Gt_*\le\Gt_+\le \pi+\Gt_*, \quad -2\pi+\Gt_*\le\Gt_-\le \Gt_*,
\label{3.3}
\eeq
where $\Gt_\pm=\arg(s\pm k)$. Then   $\Gg(0)=-ik$ and $\R \Gg(s)>0$ on $L=(-\infty+i\Gk_0,\infty+i\Gk_0)$, 
$\Gk_0\in(-k_2\sin\Gt_0,k_2)$.  On disregarding the solution 
exponentially growing at infinity we write the general solution of equation (\ref{3.2})
\beq
\GF_{0}(s,y)=A_0(s)e^{\Gg y}, \quad y<0, \quad \GF_{1}(s,y)=A_1(s)e^{-\Gg y}, \quad y>0.
\label{3.4}
\eeq
Applying the Laplace transforms (\ref{3.1}) to the boundary conditions (\ref{2.11}) gives
four equations. They are
$$
\fr{d}{dy}\GF_{0-}(s;0)=0, \quad \fr{d}{dy}\GF_{1-}(s;0)=0,
$$$$
\fr{d}{dy}\GF_{1+}(s;0)-\fr{d}{dy}\GF_{0+}(s;0)+k\tau \GF_{1+}(s;0)=0,
$$
\beq
(\mu^2-s^2)\fr{d}{dy}\GF_{0+}(s;0)-\Ga\GF_{1+}(s;0)+\Ga\GF_{0+}(s;0)=N-\fr{2\Ga i}{s+k\sin\Gt_0}.
\label{3.5}
\eeq
Here, we used the fact that, owing to the continuity of the displacements at the point $x=0$,
\beq
\lim_{x\to 0^+}\fr{d}{dy}\Gf_{0}(x;0)=0, 
\label{3.6}
\eeq
and denoted $N=\fr{d^2}{dx dy}\Gf_{0}(0^+;0)$ ($N$ is a free constant at this stage).
On fixing $y=0$ in the general solutions defined by (\ref{3.4}) and their derivatives 
and also employing the first two equations in the system (\ref{3.6}) we find
\beq
\GF_{j+}(s;0)+\GF_{j-}(s;0)=A_j(s), \quad A_j(s)=\fr{(-1)^j}{\Gg(s)}\fr{d}{dy}\GF_{j+}(s;0).
\label{3.7}
\eeq

Analysis at infinity of the fourth equation in (\ref{3.5}) and  equations (\ref{3.7}) shows that
\beq
\fr{d}{dy}\GF_{0+}(s;0)=O(s^{-2}), \quad A_0(s)=O(s^{-3}), \quad A_1(s)=O(s^{-2}), \quad s\to\infty.
\label{3.7'}
\eeq
As for the asymptotics at infinity of the other functions in the system (\ref{3.5}), they ensue from the
Laplace representations (\ref{3.1}) and the asymptotics (\ref{3.7'}). We have
$$
\GF_{j\pm}(s;0)=O(s^{-1}), \quad s\in{\Bbb C^\pm}, \quad j=0,1, \quad \fr{d}{dy}\GF_{1+}(s;0)=O(s^{-1}), \quad s\in{\Bbb C^+},
\quad s\to\infty, 
$$
\beq
 \GF_{0+}(s;0)+ \GF_{0-}(s;0)=O(s^{-3}), \quad  \GF_{1+}(s;0)+ \GF_{1-}(s;0)=O(s^{-2}), \quad |s|\to \infty, \quad s\in L.
 \label{3.7''}
 \eeq

Now, the relations (\ref{3.7}) enable us to eliminate the derivatives $\fr{d}{dy}\GF_{j+}(s;0)$ from the third and fourth equations of the
system (\ref{3.5})  and obtain
$$
[(\mu^2-s^2)\Gg(s)+\Ga]\GF_{0+}(s;0)-\Ga\GF_{1+}(s;0)+(\mu^2-s^2)\Gg(s)\GF_{0-}(s;0)=N-\fr{2\Ga i}{s+k\sin\Gt_0},
$$
\beq
-\Gg(s)\GF_{0+}(s;0)+(k\tau-\Gg)\GF_{1+}(s;0)-\Gg(s)\GF_{0-}(s;0)-\Gg(s)\GF_{1-}(s;0)=0.
\label{3.8}
\eeq
To rewrite this system in the matrix-vector form, we denote $\GF_j^\pm(s)=\GF_{j\pm}(s,0)$ and introduce the vectors
$$
\GF^\pm(s)=\left(\begin{array}{c}
\GF_0^\pm(s)\\
\GF_1^\pm(s)\\
\end{array}\right),
$$
\beq
g(s)=\fr{1}{\Gg(s)(s^2-\mu^2)}\left(N-\fr{2\Ga i}{s+k\sin\Gt_0}\right)\left(\begin{array}{c}
-1 \\
1
\end{array}\right),
\label{3.9}
\eeq
and the matrix 
\beq
G(s)=
\left(\begin{array}{cc}
b(s)+c(s)l(s) & c(s) m \\
c(s)m & b(s)-c(s)l(s)
\end{array}\right),
\label{3.10}
\eeq
where $b(s)$ and $c(s)$ are H\"older functions on the contour $L$, $l(s)$ is a polynomial and $m$ is a constant given by
$$
b(s)=1-\fr{1}{\Gg(s)}\left(\fr{k\tau}{2}+\fr{\Ga}{s^2-\mu^2}\right),\quad 
c(s)=\fr{k\tau}{2\Gg(s)(s^2-\mu^2)},
$$
\beq
l(s)=s^2-\mu^2, \quad m=\fr{2\Ga}{k\tau}.
\label{3.11}
\eeq
In these notations, the system  (\ref{3.8})  can be reformulated as the following vector Riemann-Hilbert problem
of the theory of analytic functions.

{\sl Find two vectors $\GF^+(s)$ and $\GF^-(s)$ analytic in the upper and lower half-planes ${\Bbb C^+}$ and    ${\Bbb C^-}$,
respectively, H\"older-continuous up to the contour $L$, such that their limit values on the boundary satisfy the vector equation
\beq
G(t)\GF^+(t)+\GF^-(t)=g(t), \quad t\in L.
\label{3.12}
\eeq
The solution vanishes at infinity, $\GF^\pm(s)=O(s^{-1})$, $s\to\infty$, $s\in {\Bbb C^\pm}$, and 
\beq
 \GF_0^+(t)+ \GF_0^-(t)=O(t^{-3}), \quad  \GF_1^+(t)+ \GF_1^-(t)=O(t^{-2}), \quad |t|\to \infty, \quad t\in L.
\label{3.13}
\eeq
Also, due to the condition (\ref{3.6}), 
\beq
\int_L \Gg(t) [\GF^+_{0}(t)+\GF^-_{0}(t)]dt=0.
\label{3.13.0}
\eeq}

\subsection{Matrix factorization}

The matrix coefficient $G(s)$ of the vector Riemann-Hilbert problem is a Chebotarev-Khrapkov matrix  {\bf(\ref{che})}, {\bf(\ref{khr})}.
Its characteristic polynomial is a degree-4 polynomial
\beq
f(s)=l^2(s)+m^2=(s^2-\mu^2)^2+m^2.
\label{3.14'}
\eeq
In this case the problem of factorization reduces  to a scalar Riemann-Hilbert problem
on a  two-sheeted genus-1   Riemann surface $\CR$ of the algebraic function  $w^2=f(s)$. 
Fix a single branch of this function by the condition $f^{1/2}(s)\sim s^2$, $s\to\infty$, in the plane $\hat{\Bbb C}$
cut along  the segments $\GG_1=[s_1, s_2]$ and $\GG_2=[-s_2,-s_1]$. Here,  $\pm s_1$ and $\pm s_2$ are the four
zeros of the function $f(s)$, $s_1^2=\mu^2+im$, $s_2^2=\mu^2-im$,
$s_1$ and $s_2$ are  the zeros lying in the second and first quadrant, respectively (Fig. 2).
\begin{figure}[t]
\centerline{
\scalebox{0.5}{\includegraphics{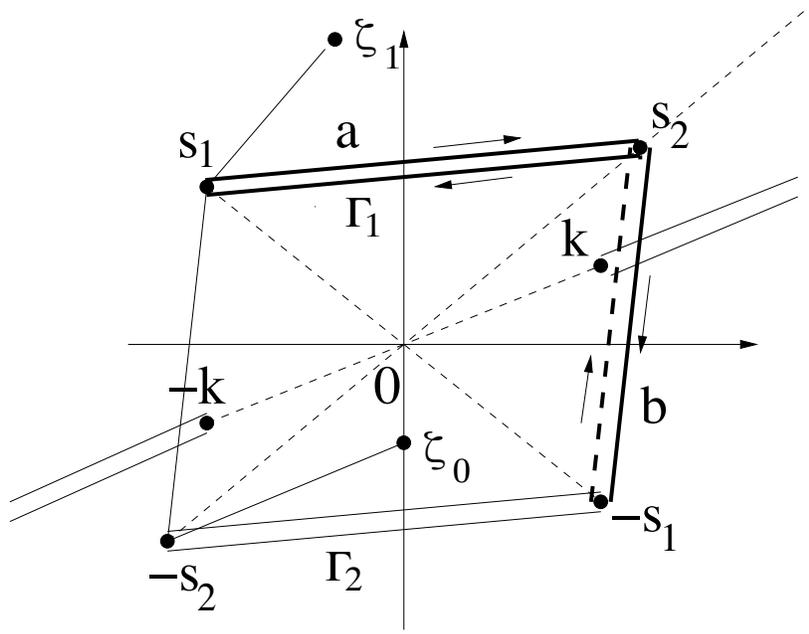}}
}
\caption{The cuts $\GG_1$ and $\GG_2$ and the canonical cross-sections { \bf a} and {\bf b}.}
\label{fig1}
\end{figure}
Denote the two sheets of the surface $\CR$  glued along the cuts $\GG_1$
and $\GG_2$  by $\Bbb C_1$ and $\Bbb C_2$.
Let $w=f^{1/2}(s)$, $(s,w)\in {\Bbb C_1}$ and $w=-f^{1/2}(s)$, $(s,w)\in {\Bbb C_2}$.

A meromorphic solution of the factorization problem 
\beq
G(t)=X^+(t)[X^-(t)]^{-1}=[X^-(t)]^{-1}X^+(t), \quad t\in L,
\label{3.14}
\eeq
the matrix $X(s)$ and its inverse, has the form  {\bf(\ref{moi})},  {\bf(\ref{am})},  {\bf(\ref{as1})}
$$
X(s)=F(s,w)B(s,w)+F(s,-w)B(s,-w),
$$
\beq
[X(s)]^{-1}=\fr{B(s,w)}{F(s,w)}+\fr{B(s,-w)}{F(s,-w)},
\label{3.15}
\eeq
where
\beq
B(s,w)=\fr12\left(I+\fr{A(s)}{w}\right), \quad A(s)=\left(\begin{array}{cc}
l(s) &  m \\
m & -l(s)
\end{array}\right).
\label{3.16}
\eeq
The function $F(s,w) $ solves the following scalar Riemann-Hilbert problem on the contour $\CL=L_1\cup L_2, $
where $L_1\subset \Bbb C_1$  and $L_2\subset \Bbb C_2$ are two copies of the contour $L$. 

{\sl Find a function $F(s,w)$  piece-wise meromorphic on the surface $\CR$,  H\"older-continuous up to the contour $\CL$,
bounded at the two infinite points of the surface $\CR$, such that
the limit values of the function $F(s,w)$  on the contour $\CL$ satisfy the relation 
\beq
F^+(t,\Gx)=\Gl(t,\Gx)F^-(t,\Gx), \quad (t,\Gx)\in \CL,
\label{3.17}
\eeq
where $\Gx=w(t)$,
$\Gl(t,\Gx)=\Gl_j(t)$ on $\Bbb C_j$, $j=1,2$, and $\Gl_1(t)=b(t)+ c(t)f^{1/2}(t)$ and $\Gl_2(t)=b(t)- c(t)f^{1/2}(t)$
are the two eigenvalues of the matrix $G(t)$.} 

It is directly verified that the increments of the arguments of the eigenvalues $\Gl_1(t)$ and $\Gl_2(t)$ when
$t$ traverses the contour $L$ from $-\infty+i\Gk_0$ to $+\infty+i\Gk_0$ are equal to zero. Therefore the general solution of the problem (\ref{3.17})
has the form
\beq
F(s,w)=e^{\Gc(s,w)}, \quad (s,w)\in\CR,
\label{3.18}
\eeq
where 
\beq
\Gc(s,w)=\fr{1}{2\pi i}\int_\CL\log\Gl(t,\Gx)dW+\int_\GG dW+m_a\int_{\rm \Ba} dW +m_b\int_{\rm \Bb} dW, 
\label{3.19}
\eeq
$dW$ is  the Weierstrass analog of the Cauchy kernel on an elliptic surface,
\beq
dW=\fr{w+\Gx}{2\Gx}\fr{dt}{t-s}
\label{3.20}
\eeq
and $\GG$ is a contour on the surface $\CR$
whose starting point $q_0$ is fixed arbitrarily say, $q_0=(\Gz_0,f^{1/2}(\Gz_0))\in{\Bbb C_1}$, while the terminal
point $q_1=(\Gz_1,w(\Gz_1))\in\CR$ cannot be fixed a priori and has to be determined. In formula (\ref{3.19}),
there are two more undetermined quantities, $m_a$ and $m_b$. They are integers and have also to be determined.
The contours of integration ${\rm \bf a}$ and ${\rm \bf b}$ are canonical cross-sections of the surface $\CR$ (Fig. 2).
The cross-section  ${\rm \bf a}={\rm \bf a^+}\cup {\rm \bf a^-}$ is a two-sided loop, ${\rm \bf a^+}\in{\Bbb C_1}$
and  ${\rm \bf a^-}\in{\Bbb C_2}$. The closed contour ${\rm \bf b}$ consists os the segment $[s_2,-s_1]\in{\Bbb C_1}$
and the segment $[-s_1,s_2]\in{\Bbb C_2}$ (the dashed line in Fig.2). The loop ${\rm \bf a}$ intersects the loop
${\rm \bf b}$ at the branch point $s_2$ from left to right. Note that the contour $\GG$ has to be chosen such that it 
intersects neither the contour ${\rm \bf a}$ no the contour  ${\rm \bf b}$.

Because of the logarithmic singularities of the second integral in (\ref{3.19}) at the endpoints of the contour $\GG$
the solution $F(s,w)$ has a simple pole at the point $q_0\in{\Bbb C_1}$ and a simple zero at the point $q_1\in\CR$.
The kernel $dW$ has an order-2 pole at  the infinite points of the surface. That is why the solution
$F(s,w)$ has unacceptable essential singularities at the infinite points. To remove them, we first rewrite $\Gc(s,w)$
in the form
\beq
\Gc(s,w)=\Gc_1(s)+w\Gc_2(s),
\label{3.21}
\eeq
where
$$
\Gc_1(s)=\fr{1}{4\pi i}\int_L\fr{\log\GD(t)dt}{t-s}+\fr12\int_\GG\fr{dt}{t-s},
$$
$$
\Gc_2(s)=\fr{1}{4\pi i}
\int_L\fr{\log\Ge(t) dt}{(t-s)f^{1/2}(t)}+\fr12\left(\int_\GG+m_a\int_{\Ba}+m_b\int_{\Bb}\right)\fr{dt}{(t-s)\Gx},
$$
\beq
\GD(t)=\Gl_1(t)\Gl_2(t), \quad \Ge(t)=\fr{\Gl_1(t)}{\Gl_2(t)}.
\label{3.22}
\eeq
The function $\Gc(s,w)$
is bounded at infinity if and only if $w\Gc_2(s)=O(1)$, $s\to \infty$, or, equivalently,
\beq
\fr{1}{2\pi i}\int_L\fr{\log\Ge(t)dt}{f^{1/2}(t)}+\int_\GG\fr{dt}{\Gx}+m_a\int_{\Ba}\fr{dt}{\Gx}+m_b\int_{\Bb}\fr{dt}{\Gx}=0.
\label{3.23}
\eeq
If the function $\Gc_2(s)$ satisfies this condition, then because of the identity 
\beq
\fr{1}{t-s}=-\fr{1}{s}+\fr{t}{s(t-s)},
\label{3.23'}
\eeq
 it admits an alternative representation
\beq
\Gc_2(s)=\fr{1}{4\pi i s} 
\int_L\fr{\log\Ge(t)  tdt}{(t-s)f^{1/2}(t)}+\fr{1}{2s}\left(\int_\GG+m_a\int_{\Ba}+m_b\int_{\Bb}\right)\fr{tdt}{(t-s)\Gx}.
\label{3.23''}
\eeq
This formula is valid for all $s\ne 0$ and can be used   for numerical calculations when  $|s|>1$
instead of formula (\ref{3.22})
that is more convenient  when $|s|\le 1$.

\subsection{Jacobi inversion problem}

The condition (\ref{3.23}) is equivalent to a genus-1 Jacobi inversion problem. To show this, consider
the abelian (elliptic) integral
\beq
\Go(q)=\int_{-s_2}^q\fr{ds}{w(s)},\quad q=(s,w(s)).
\label{3.24}
\eeq
Then the third and fourth integrals in equation (\ref{3.23}),
$$
\CA=\int_{\Ba}\fr{ds}{w(s)}=2\int_{\Ba^+}\fr{ds}{f^{1/2}(s)}, 
$$
\beq
\CB=\int_{\Bb}\fr{ds}{w(s)}=2\int_{s_2}^{-s_1}\fr{ds}{f^{1/2}(s)},
\label{3.25}
\eeq
are the $A$- and $B$-periods of the integral $\Go(q)$.
Thus the condition  (\ref{3.23}) constitute the following Jacobi inversion problem.

{\sl  Find a point $q_1=(\Gz_1,w_1)\in\CR$ and two integers $m_a$ and $m_b$ such that
\beq
\Go(q_1)+m_a\CA+m_b\CB=d_0,
\label{3.26}
\eeq
where
\beq
d_0=\Go(q_0)-\fr{1}{2\pi i}\int_L\fr{\log\Ge(t)dt}{f^{1/2}(t)}.
\label{3.27}
\eeq}

By dividing equation (\ref{3.26}) by $\CA$ we arrive at the canonical form of the Jacobi problem 
\beq
\hat\Go(q_1)=e_1-k_1-m_a-m_b\hat\CB\equiv e_1-k_1 \quad (\rm{modulo \;the\; periods})
\label{3.28'}
\eeq
for the canonical abelian integral  $\hat\Go(q)=\fr{\Go(q)}{\CA}$. It has the unit $A$-period, while its $B$-period  
$\hat\CB=\fr{\CB}{\CA}$  has a
positive imaginary part,  $\I\hat\CB>0$. Here,
$e_1=\fr{d_0}{\CA}+k_1$ and 
$k_1$ is the Riemann constant of the surface $\tilde\CR$ cut along the loops $\rm \Ba$ and $\rm \Bb$. It is 
computed in {\bf(\ref{as2})}, $k_1=-\fr12+\fr12\hat\CB$. 

The unknown point $q_1=(\Gz_1,w_1)$ is the single zero of the genus-1 Riemann $\Gt$-function
\beq
\CF(q)=\Gt(\hat\Go(q)-e_1)=\sum_{\nu=-\infty}^\infty \exp\{\pi i\hat\CB \nu^2+2\pi i \nu[\hat\Go(q)-e_1]\}.
\label{3.28}
\eeq
To find $\Gz_1$, consider the integral 
\beq
M=\fr{1}{2\pi i} \int_{\Md\hat\CR}\fr{d\log\CF(q)}{s-i},
\label{3.29}
\eeq
where $\Md\hat\CR$ is the boundary of the surface $\CR$ cut along the cross-section ${\Ba}$ only.
The procedure we are going to apply is a modification of the method  {\bf(\ref{khr})} that instead of the poles
at the two points $i_1=(i,f^{1/2}(i))\in{\Bbb C_1}$ and $i_2=(i,-f^{1/2}(i))\in{\Bbb C_2}$ in (\ref{3.29}) 
uses an integrand with two poles at the infinite points of the surface $\CR$.
Without loss we assume that $\CF(i_n)\ne 0$, $n=1,2$, and compute the integral $M$ by applying the theory 
of residues. We have
\beq
M=\fr{1}{\Gz_1-i}+\fr{\CF'(i_1)}{\CF(i_1)}+\fr{\CF'(i_2)}{\CF(i_2)}.
\label{3.30}
\eeq
where the derivative of the  Riemann $\Gt$-function is a rapidly convergent series
\beq
\CF'(q)=\fr{2\pi i (-1)^{n-1}}{\CA f^{1/2}(s)}\sum_{\nu=-\infty}^\infty \nu \exp\{\pi i\hat\CB \nu^2+2\pi i \nu[\hat\Go(q)-e_1]\}, 
\quad q\in{\Bbb C_n}, \quad n=1,2. 
\label{3.30'}
\eeq
The  integral (\ref{3.29}) can also be represented as a contour integral
\beq
M=\fr{1}{2\pi i}\left(\int_{\Ba^+}+\int_{\Ba^-}\right)\fr{d\log\CF(q)}{s-i}
=\fr{1}{2\pi i}\int_{\Ba} \fr{1}{s-i}[d\log \CF^+(q)-d\log\CF^-(q)],
\label{3.31}
\eeq
where $\Ba^+$ and $\Ba^-$ have opposite directions and pointwise coincide with the loop $\Ba$ from the side of the sheets $\Bbb C_1$ and  $\Bbb C_2$, respectively, while  the positive direction of the loops $\Ba^+$ and $\Ba$
are chosen such that the exterior of the cut $[s_1,s_2]$ on the first sheet is on the left.
Using the relation between the boundary values of the Riemann $\Gt$-function
on the loop $\Ba$
\beq
\CF^+(p)=\CF^-(p)\exp\{\pi i\hat\CB-2\pi i e_1+2\pi i\hat\Go^+(p)\}, \quad p\in {\rm \Ba},
\label{3.32}
\eeq
where
\beq
\hat\Go^+(p)=\lim_{q\to p, q\in{\Bbb C_1}}\hat\Go(q),
\label{3.33}
\eeq
we derive another formula for the integral $M$
\beq
M=\int_{\Ba} \fr{d\hat\Go^+(q)}{s-i}=\fr{2}{\CA}\int_{[s_1,s_2]^+}\fr{ds}{(s-i)f^{1/2}(s)}.
\label{3.34}
\eeq
Combining formulas (\ref{3.30}) and (\ref{3.34}) and introducing the quantity
\beq
\CJ=\fr{2}{\CA}\int_{[s_1,s_2]^+}\fr{ds}{(s-i)f^{1/2}(s)}-\fr{\CF'(i_1)}{\CF(i_1)}-\fr{\CF'(i_2)}{\CF(i_2)}
\label{3.34'}
\eeq
we find the parameter $\Gz_1$ 
\beq
\Gz_1=i+\fr{1}{\CJ}.
\label{3.35}
\eeq
Evaluate now the abelian integral at the point $\Gz_1$ lying on the first and second sheets,
\beq
\hat\Go(\Gz_1,\pm f^{1/2}(\Gz_1))=\pm\fr{1}{\CA}
\int_{-s_2}^{\Gz_1}\fr{ds}{f^{1/2}(s)}.
\label{3.36}
\eeq
and denote
\beq
d_\pm=\fr{d_0}{\CA}-\hat\Go(\Gz_1,\pm f^{1/2}(\Gz_1)).
\label{3.37}
\eeq
Taking the imaginary and real parts of the complex equation (\ref{3.32}) we determine
the constants $m^\pm_b$ and $m^\pm_a$
\beq
m^\pm_b=\fr{\I d_\pm}{\I\hat\CB}, \quad m^\pm_a=\R d_\pm-\fr{\R\hat\CB}{\I\hat\CB}\I d_\pm.
\label{3.38}
\eeq
If it turns out that both of the integers $m^+_a$ and $m_b^+$ are integers, then $m_a=m_a^+$,
 $m_b=m_b^+$, and the point $q_1\in {\Bbb C_1}$. Otherwise,    $m_a=m_a^-$,
 $m_b=m_b^-$ are integers, and  $q_1\in {\Bbb C_2}$.

\subsection{Solution of the vector Riemann-Hilbert problem}

On having factorized the matrix $G(t)=X^+(t)[X^-(t)]^{-1}=[X^-(t)]^{-1}X^+(t)$, $t\in L$, and eliminated the essential singularity
of the factors $X^\pm(s)$ at infinity, we proceed with the solution of the vector Riemann-Hilbert problem (\ref{3.12})
by rewriting the boundary relation as
\beq
X^+(t)\GF^+(t)-N\GY_+^{(1)}(t)-\GY_+^{(2)}(t)=-X^-(t)\GF^-(t)-N\GY_-^{(1)}(t)-\GY_-^{(2)}(t), \quad t\in L.
\label{3.39}
\eeq
Here, $\GY^{(n)}_\pm(t)$ are the limit values of the Cauchy integrals 
$$
\GY^{(1)}(s)=\fr{1}{2\pi i}\int_L\fr{X^-(t)J}{\Gg(t)(t^2-\mu^2)}\fr{dt}{t-s},\quad
 J=\left(\begin{array}{c}
-1 \\
1
\end{array}\right),
$$
\beq
\GY^{(2)}(s)=-\fr{\Ga}{\pi}\int_L\fr{X^-(t)J}{\Gg(t)(t^2-\mu^2)(t+k\sin\Gt_0)}\fr{dt}{t-s},
\label{3.39'}
\eeq
defined by the Sokhotski-Plemelj formulas
$$
\GY_\pm^{(1)}(t)=\pm\fr{X^-(t)J}{2\Gg(t)(t^2-\mu^2)}+P.V. \GY^{(1)}(t),
$$
\beq
\GY_\pm^{(2)}(t)=\mp\fr{\Ga i X^-(t)J}{\Gg(t)(t^2-\mu^2)(t+k\sin\Gt_0)}+P.V. \GY^{(2)}(t), \quad t\in L.
\label{3.40}
\eeq
By the continuity principle and the generalized Liouville's theorem of the theory of analytic functions, 
\beq
\pm X^\pm(s)\GF^\pm(s)-N \GY_\pm^{(1)}(s)-\GY_\pm^{(2)}(s)=R(s), \quad s\in{\Bbb C^\pm},
\label{3.41}
\eeq
where $R(s)$ is a rational vector-function to be determined.
Since the matrices $X^\pm(s)$ are bounded at infinity, while the vectors $\GF^\pm(s)$, $\GY_\pm^{(1)}(s)$ and
$ \GY_\pm^{(2)}(s)$  behave as $s^{-1}\const $ for large $s$, we have $R(s)=O(s^{-1})$,  $s\to\infty$.

Show next that  the rational vector-function $R(s)$ has  a pole at the point $s=\Gz_0$.
Indeed, due to the logarithmic singularity of the function $\Gc(s,w)$
  at the starting point $(\Gz_0,f^{1/2}(\Gz_0))\in{\Bbb C_1}$ of the contour $\GG$, the function $F(s,w)$ has a simple pole at this point
  and is bounded at the  point  $(\Gz_0,-f^{1/2}(\Gz_0))\in{\Bbb C_2}$,
  \beq
  F(s,w)\sim D_0(s-\Gz_0)^{-1}, \quad F(s,-w)=O(1), \quad s\to\Gz_0, \quad (s,w)\in{\Bbb C_1}.
  \label{3.42}
  \eeq
   Employing the first formula in (\ref{3.15})
for the matrix $X(s)$ we obtain
\beq
X(s)\GF(s)\sim \fr{D_0}{2(s-\Gz_0)}\left[\left(1+\fr{l(\Gz_0)}{f^{1/2}(\Gz_0)}\right)\GF_0(\Gz_0)+\fr{m}{f^{1/2}(\Gz_0)}
\GF_1(\Gz_0)\right]\left(\begin{array}{c}
1 \\
\Gn_0
\end{array}\right),
\label{3.43}
\eeq
where
\beq
\Gn_0=\fr{m}{l(\Gz_0)+f^{1/2}(\Gz_0)}.
\label{3.44}
\eeq
This enables us to find the vector $R(s)$. We have
\beq
R(s)=\fr{C}{s-\Gz_0}\left(\begin{array}{c}
1 \\
\Gn_0
\end{array}\right),
\label{3.45}
\eeq
where $C$ is a free constant. By substituting this expression into equation (\ref{3.41}) we get the solution
 of the  vector Riemann-Hilbert problem (\ref{3.12}) 
\beq
\GF^\pm(s)=\pm [X^\pm(s)]^{-1}\left[\fr{C}{s-\Gz_0}\left(\begin{array}{c}
1 \\
\Gn_0
\end{array}\right)+N \GY_\pm^{(1)}(s)+\GY_\pm^{(2)}(s)\right], \quad s\in{\Bbb C^\pm}.
\label{3.46'}
\eeq
Now, the function $F(s,w)$ has a simple zero at the point $q_1\in\CR$ caused by the logarithmic singularity 
of the function $\Gc(s,w)$ at the terminal point of the contour $\GG$. If $q_1=(\Gz_1,w_1)\in{\Bbb C_1}$,
then $w_1=f^{1/2}(\Gz_1)$ and
\beq
F(s,w)\sim D_1(s-\Gz_1), \quad F(s,-w)\sim D_2, \quad (s,w)\in{\Bbb C_1}, \quad s\to\Gz_1.
\label{3.46}
\eeq
Otherwise, is $q_1=(\Gz_1,w_1)\in{\Bbb C_2}$,
then $w_1=-f^{1/2}(\Gz_1)$ and
\beq
F(s,w)\sim D_2, \quad F(s,-w)\sim D_1(s-\Gz_1), \quad (s,w)\in{\Bbb C_1}, \quad s\to\Gz_1.
\label{3.47}
\eeq
Here,  $D_1$ and $D_2$ are  nonzero constants. From the second formula in (\ref{3.15}) 
for the inverse matrix $[X(s)]^{-1}$, it becomes evident that the matrix $[X(s)]^{-1}$ has a pole at the point
$s=\Gz_1$. Since ${\rm rank} \, B(s,w)=1$, the vector-function $F(s,w)$ has a removable singularity at the point $s=\Gz_1$
if and only if 
\beq
\GP_0 C+\GP_1 N=-\GP_2,
\label{3.48}
\eeq
where
$$
\GP_0=\left(\fr{w_1+l(\Gz_1)}{m}+\Gn_0\right)\fr{1}{\Gz_1-\Gz_0},\quad
\GP_1=\fr{w_1+l(\Gz_1)}{m}\GY_0^{(1)}(\Gz_1)+\GY_1^{(1)}(\Gz_1),
$$
\beq
\GP_2=\fr{w_1+l(\Gz_1)}{m}\GY_0^{(2)}(\Gz_1)+\GY_1^{(2)}(\Gz_1),
\label{3.49}
\eeq
and $\GY_0^{(n)}(s)$ and $\GY_1^{(n)}(s)$ ($n=1,2$) are the two components of the vectors  $\GY^{(n)}(s)$
given by (\ref{3.39'}). Equation (\ref{3.48}) constitutes the first equation for the unknown constants $C$ and $N$.

The solution derived has to be restricted to the class of functions satisfying the conditions (\ref{3.13}). 
To verify these conditions, we examine the asymptotics of the vectors $\GF^+(t)$ and $\GF^-(t)$ as $|t|\to\infty$, $t\in L$.
We start with the analysis of the functions $\GD(t)$ and $\Ge(t)$. From formulas (\ref{3.11}) and (\ref{3.22}) we have
$$
\GD(t)=1-\fr{k\tau}{\Gg(t)}-\fr{2\Ga}{\Gg(t)(t^2-\mu^2)}+\fr{k\tau\Ga}{\Gg^2(t)(t^2-\mu^2)}= 1-\fr{k\tau}{|t|}+O(t^{-3}),\quad |t|\to\infty,
\quad t\in L,
$$
\beq
\Ge(t)=\fr{b(t)+c(t)f^{1/2}(t)}{b(t)-c(t)f^{1/2}(t)}=1+\fr{k\tau}{|t|}+O(t^{-2}),\quad |t|\to\infty, \quad t\in L,
\label{3.50}
\eeq
By applying the Sokhotski-Plemelj formulas to the integral representations of the functions $\Gc_1(s)$ and $\Gc_2(s)$
given by (\ref{3.22}) and (\ref{3.23''}) we deduce
$$
\Gc_1^\pm(t)=\pm\fr{\log\GD(t)}{4}+\fr{1}{4\pi i}
\int_L\fr{\log\GD(t_0) dt_0}{t_0-t}+
\fr12\int_\GG
\fr{dt_0}{t_0-t},\quad t\in L,
$$
$$
f^{1/2}(t)\Gc^\pm_2(t)=\pm\fr{\log\Ge(t)}{4}+
\fr{f^{1/2}(t)}{4\pi i t} 
\int_L\fr{\log\Ge(t_0)  t_0dt_0}{(t_0-t)f^{1/2}(t_0)}
$$
\beq
+\fr{f^{1/2}(t)}{2t}\left(\int_\GG+m_a\int_{\Ba}+m_b\int_{\Bb}\right)\fr{t_0dt_0}{(t_0-t)
\Gx(t_0)}, \quad t\in L\setminus\{0\}.
\label{3.51}
\eeq
We focus our attention  on the principal terms of the asymptotic expansions at infinity of the two functions in (\ref{3.51}) 
and observe that
$$
\Gc_1^\pm(t)=\mp\fr{k\tau}{4|t|}+\fr{k\tau}{2\pi i t}\log|t|+\fr{h_1}{t}+O(t^{-2}), \quad |t|\to\infty, \quad t\in L,
$$
\beq
f^{1/2}(t)\Gc^\pm_2(t)=h_0\pm\fr{k\tau}{4|t|}-\fr{k\tau}{2\pi i t}\log|t|+\fr{h_2}{t}+O(t^{-2}), \quad |t|\to\infty, \quad t\in L,
\label{3.52}
\eeq
where
\beq
h_0=-\fr12\left(\int_\GG+m_a\int_{\Ba}+m_b\int_{\Bb}\right)\fr{tdt}{\Gx(t)}
\label{3.53}
\eeq
and $h_1$ and $h_2$ are some constants. Their values do not affect the asymptotics we aim to derive.

Next, by virtue of the relation (\ref{3.21}) we can derive the asymptotics of the solution of the Riemann-Hilbert problem
on the surface $\CR$ as follows:
$$
F^\pm(t,f^{1/2}(t))=e^{h_0}\left[1+\fr{h_+}{t}+O(t^{-2})\right], \quad  |t|\to\infty, \quad t\in L,
$$
\beq
F^\pm(t,-f^{1/2}(t))=e^{-h_0}\left[1\mp\fr{k\tau}{2|t|}+\fr{k\tau}{\pi i t}\log|t|+\fr{h_-}{t}+O\left(\fr{\log^2|t|}{t^2}\right)\right], \quad  
|t|\to\infty, \quad t\in L,
\label{3.54}
\eeq
where $h_\pm=h_1\pm h_2$. Substituting these expressions into formula (\ref{3.15}) for the inverse matrix $[X(s)]^{-1}$
gives 
$$
[X^\pm(t)]^{-1}=\left(\begin{array}{cc}
e^{-h_0}(1-h_+t^{-1}) &  0 \\
0 & e^{h_0}[1-k\tau(\pi i t)^{-1}\log|t|-h_-t^{-1}]+O(t^{-2}\log^2|t|)
\end{array}\right)
$$
\beq
\pm\fr{k\tau e^{h_0}}{2|t|}
\left(\begin{array}{cc}
0 &  0 \\
0 & 1
\end{array}\right)+O\left(\fr{1}{t^2}\right), \quad |t|\to\infty, \quad t\in L,
\label{3.55}
\eeq
The above result, together with formulas (\ref{3.40}) and (\ref{3.46'}) enables us to derive formulas which describe
the behavior of the solution $\GF^\pm(t)$ of the vector Riemann-Hilbert problem on the contour $L$ when $|t|\to\infty$
$$
\GF^\pm(t)=
\pm\fr{C_0}{t}\left(\begin{array}{cc}
e^{-h_0}(1-h_+t^{-1}) &  0 \\
0 & e^{h_0}[1-k\tau(\pi i t)^{-1}\log|t|-h_-t^{-1}]+O(t^{-2}\log^2|t|)
\end{array}\right)
$$
\beq
+\fr{C_0k\tau e^{h_0}}{2|t|t}
\left(\begin{array}{cc}
0 &  0 \\
0 & 1
\end{array}\right)
+O\left(\fr{1}{t^3}\right), \quad |t|\to\infty, \quad t\in L,
\label{3.56}
\eeq
where $C_0$ is a constant. From here we immediately get $\GF^+_0(t)+\GF^-_0(t)=O(t^{-3})$ and  $\GF^+_1(t)+\GF^-_1(t)=O(t^{-2})$,
$|t|\to\infty$, $t\in L$, and the conditions (\ref{3.13}) are  fulfilled as required.

Finally, we satisfy the condition (\ref{3.13.0}) that guarantees the continuity of the displacement at the junction point $x=0$.
The asymptotics of the sum $\GF^+_0(t)+\GF^-_0(t)=O(t^{-3})$, $t\to\infty$, $t\in L$, we just verified, is sufficient for the convergence of the 
integral in  (\ref{3.13.0}). To transform the condition  (\ref{3.13.0})  into an equation with respect to the constants $C$ and $N$, we
recast the formula (\ref{3.15}) and express $[X^\pm(t)]^{-1}$ through functions on the complex plane
\beq
[X^\pm(t)]^{-1}=e^{-\Gc_1^\pm(t)}\left\{I\cosh[f^{1/2}(t)\Gc_2^\pm(t)]
- \left(\begin{array}{cc}
t^2-\mu^2  &  m \\
m & -t^2+\mu^2
\end{array}\right)
\fr{\sinh[f^{1/2}(t)\Gc_2^\pm(t)]}
{f^{1/2}(t)}
\right\}.
\label{3.57}
\eeq
It is convenient to introduce the functions
$$
\GL_j^\pm(t)=e^{-\Gc_1^\pm(t)}[\cosh(f^{1/2}(t)\Gc_2^\pm(t))+\fr{t^2-\mu^2}{(-1)^j f^{1/2}(t)}\sinh(f^{1/2}(t)\Gc_2^\pm(t))],\quad j=0,1,
$$
\beq
\GL_2^\pm(t)=\fr{me^{-\Gc_1^\pm(t)}}{f^{1/2}(t)}\sinh(f^{1/2}(t)\Gc_2^\pm(t)).
\label{3.58}
\eeq
In terms of these functions, the functions $\GF_0^\pm(t)$ can be written as follows:
$$
\pm\GF_0^\pm(t)=\GL_1^\pm(t)\left[\fr{C}{t-\Gz_0}+N\GY_{0\pm}^{(1)}(t)+
\GY_{0\pm}^{(2)}(t)\right]
$$
\beq-\GL_2^\pm(t)\left[\fr{C\Gn_0}{t-\Gz_0}+N\GY_{1\pm}^{(1)}(t)+
\GY_{1\pm}^{(2)}(t)\right], \quad t\in L,
\label{3.59}
\eeq
where $\GY_{0\pm}^{(n)}(t)$ and $\GY_{1\pm}^{(n)}(t)$ are the two components of the vectors $\GY_{\pm}^{(n)}(t)$, $n=1,2$,
given by (\ref{3.40}).
On substituting these expressions into the relation  (\ref{3.13.0}) we obtain the second equation for the constants $C$ and $N$
\beq
\GO_0 C+\GO_1N=-\GO_2,
\label{3.60}
\eeq
where
$$
\GO_0=\int_L [\GL_1^+(t)-\GL_1^-(t)-\Gn_0\GL_2^+(t)+\Gn_0\GL_2^-(t)]\fr{\Gg(t)dt}{t-\Gz_0},
$$
\beq
\GO_j=\int_L [\GL_1^+(t)\GY_{0+}^{(j)}(t)-\GL_{1}^-(t)\GY_{0-}^{(j)}(t)-\GL_2^+(t)\GY_{1+}^{(j)}(t)+\GL_2^-(t)\GY_{1-}^{(j)}(t)]\Gg(t)dt,\quad j=1,2.
\label{3.61}
\eeq
The system (\ref{3.48}), (\ref{3.61}) determines the constants $C$
 and $N$
\beq
C=\fr{\GP_1\GO_2-\GP_2\GO_1}{\GD_0}, \quad N=\fr{\GP_2\GO_0-\GP_0\GO_2}{\GD_0}, 
\label{3.62}
\eeq
where $\GD_0=\GP_0\GO_1-\GP_1\GO_0.$ The determination of these constants completes the solution of the 
vector Riemann-Hilbert problem (\ref{3.12}) to (\ref{3.13.0}).

\subsection{Partial indices of factorization}

In this section we wish to determine the partial indices of factorization defined as
 the orders of the columns of the canonical matrix of factorization {\bf(\ref{gak})},  {\bf(\ref{mus})},  {\bf(\ref{vek})}
 by the method {\bf(\ref{gak})} applied in the genus 3 case {\bf(\ref{as1})}.
The canonical matrix is a matrix $X(s)$ that solves the factorization problem $X^+(t)=G(t)X^-(t)$, $t\in L$, and satisfies the conditions

(i) at any finite point $s\in{\Bbb C}$,   $X(s)$ is in normal form,
 
(ii) $\det  X(s)$ does not have zeros at any finite point in the complex plane, and 

(iii) the matrix $X(s)$ is in normal form at infinity.

We recall that a matrix $Y(s)$ is in normal form at a point (finite or infinite) if the order of the determinant of the matrix
at this point is equal to the sum of the orders of the matrix columns.

Assume $Y_j(s)=Y_j^*(s) (s-s_0)^{\Ga_j}$, $s\to s_0$, $j=1,\ldots,n$, where $\Ga_j$ is real, $Y_j^*(s)$ is bounded at 
$s=s_0$ and  $Y_j^*(s_0)\ne 0$. Then $\Ga_j$ is called the  order of the function $Y_j(s)$ at $s=s_0$ and
$\Ga=\min\{\Ga_1,\ldots, \Ga_n\}$ is called the order of the vector $Y(s)=(Y_1(s),\ldots,Y_n(s))$ at the point $s=s_0$.

Let $Y_j(s)=Y_j^*(s) s^{-\Ga_j}$, $s\to\infty$, $j=1,\ldots,n$, where $\Ga_j$ is real, $Y_j^*(s)$ is bounded at infinity and  $Y_j^*(\infty)\ne 0$. Then $\Ga_j$ and $\Ga=\min\{\Ga_1,\ldots, \Ga_n\}$ are called the order  at infinity of the function $Y_j(s)$ and  vector $Y(s)$,
respectively.

Since the properties of the inverse matrix $[X(s)]^{-1}$ have been studied in Section 3.4, we shall convert the 
matrix $[X(s)]^{-1}$, not the matrix $X(s)$ itself, into a canonical matrix.
The matrix $[X(s)]^{-1}$ has three singular points, $\Gz_0$, $\Gz_1$, and $\infty$.
At the point $s=\Gz_0$, it admits the representation
\beq
[X(s)]^{-1}\sim F_0
\left(\begin{array}{cc}
1-\fr{l_0}{w_0} &  -\fr{m}{w_0} \\
 -\fr{m}{w_0} & 1+\fr{l_0}{w_0} 
\end{array}\right)
+(s-\Gz_0)\left(\begin{array}{cc}
F_{11} &  F_{12} \\
F_{21} & F_{22}\end{array}\right), \quad  s\to \Gz_0,
\label{3.63}
\eeq
where $l_0=l(\Gz_0)$, $w_0=f^{1/2}(\Gz_0)$, and $F_0$ and $F_{n j}$ are nonzero constants.  It is clear that $\det\{[X(s)]^{-1}\}\sim 0$ as
$s\to \Gz_0$, and the order of the determinant at the point $\Gz_0$ is equal to 1, while  both columns have zero-orders.
To transform the matrix into normal form, we multiply it from the right by the matrix
\beq
 T_0(s)=
\left(\begin{array}{cc}
\fr{1}{s-\Gz_0} &  0 \\
\fr{\nu_0}{s-\Gz_0} &  1\\
\end{array}\right), \quad \nu_0=\fr{w_0-l_0}{m}.
\label{3.64}
\eeq 
The new matrix $[X(s)]^{-1}T_0(s)$ is in normal form at the point $s=\Gz_0$,
\beq
[X(s)]^{-1}T_0(s)\sim\left(\begin{array}{cc}
 F_{11}+\nu_0F_{12} &  -\fr{m}{w_0}F_0 \\
F_{21}+\nu_0F_{22}  & \left(1+\fr{l_0}{w_0}\right)F_0\\ 
\end{array}\right),\quad s\to\Gz_0.
\label{3.65}
\eeq
Proceed now with converting the new matrix into normal form at the point $\Gz=\Gz_1$. The original matrix $[X(s)]^{-1}$
behaves at the point $(\Gz_1, w_1)$ as
\beq
[X(s)]^{-1}\sim \fr{F_1}{s-\Gz_1}
\left(\begin{array}{cc}
1+\fr{l_1}{w_1} &  \fr{m}{w_1} \\
 \fr{m}{w_1} & 1-\fr{l_1}{w_1} 
\end{array}\right)
+\left(\begin{array}{cc}
\hat F_{11} &  \hat F_{12} \\
\hat F_{21} & \hat F_{22}\end{array}\right), \quad  (s,w)\to (\Gz_1,w_1),
\label{3.66}
\eeq
where  $w_1=(-1)^{n-1}f^{1/2}(\Gz_1)$,   $n=1$  if the point $ (\Gz_1,w_1)$ lies on the first sheet ${\Bbb C_1}$ and $n=2$ otherwise,  $l_1=l(\Gz_1)$, and
$F_1$ and $\hat F_{n j}$ are nonzero constants.  By multiplying the new matrix $[X(s)]^{-1}T_0(s)$  from the right  by the matrix
\beq
 T_1(s)=
\left(\begin{array}{cc} 
\Gz_1-\Gz_0 &  0 \\
\nu_1-\nu_0 &  s-\Gz_1\\
\end{array}\right), \quad \nu_1=-\fr{w_1+l_1}{m},
\label{3.67}
\eeq 
we obtain the matrix $[X(s)]^{-1}\hat T(s)$, where
\beq
\hat T(s)=T_0(s)T_1(s)=
\left(\begin{array}{cc}
\fr{\Gz_1-\Gz_0}{s-\Gz_0} &  0 \\
\nu_1-\nu_0+\fr{\nu_0(\Gz_1-\Gz_0)}{s-\Gz_0} &  s-\Gz_1\\
\end{array}\right).
\label{3.68}
\eeq 
It is directly verified that the matrix $[X(s)]^{-1}\hat T(s)$ is in normal form at both points, $s=\Gz_0$ and $\Gz_1$. 
At the point $(s,w)=(\Gz_1,w_1)$, we have
\beq
[X(s)]^{-1}\hat T(s)\sim\left(\begin{array}{cc}
\hat F_{11}+\nu_1\hat F_{12} &  \fr{m}{w_1}F_1 \\
\hat F_{21}+\nu_1\hat F_{22}  & \left(1-\fr{l_1}{w_1}\right)F_1\\ 
\end{array}\right),\quad s\to\Gz_1.
\label{3.69}
\eeq
A similar remedy is to be used to make the matrix $[X(s)]^{-1}\hat T(s)$ in normal form at infinity. Analysis of the matrix
$[X(s)]^{-1}\hat T(s)$ at infinity shows
\beq
[X(s)]^{-1}\hat T(s)\sim\left(\begin{array}{cc}
e^{-h_0}(\Gz_1-\Gz_0) s^{-1} &  h_{12}s^{-1} \\
e^{h_0}(\nu_1-\nu_0)  &e^{h_0}s\\ 
\end{array}\right),\quad s\to\infty,
\label{3.70}
\eeq
where $h_0$ is given by (\ref{3.53}) and $h_{12}$ is a nonzero constant.
The orders of the columns of the matrix $[X(s)]^{-1}\hat T(s)$ at infinity are equal to 0 and $-1$, and their sum does not equal
the order 0 of the determinant of this matrix at infinity. We multiply it from the right by the matrix
\beq
 U(s)=
\left(\begin{array}{cc}
0 &  \nu s \\
0 &  1 \\
\end{array}\right), 
\label{3.71}
\eeq 
and arrive at the matrix $\tilde X(s)=[X(s)]^{-1}T(s)$ that is in normal form at the infinite point if the parameter $\nu$ is chosen to be
$\nu=(\nu_0-\nu_1)^{-1}$. Then
\beq
\tilde X(s)\sim \left(\begin{array}{cc}
e^{-d_0}(\Gz_1-\Gz_0) s^{-1} &  e^{-d_0}\fr{\Gz_1-\Gz_0}{\nu_0-\nu_1} \\
e^{d_0}(\nu_1-\nu_0)  & d_{22}\\ 
\end{array}\right),\quad s\to\infty,
\label{3.72}
\eeq
where $d_{22}$ is a constant.   The transformation matrix  $T(s)=\hat T(s)U(s)$ is
\beq
T(s)=
\left(\begin{array}{cc}
\fr{\Gz_1-\Gz_0}{s-\Gz_0} &  \fr{(\Gz_1-\Gz_0)s}{(s-\Gz_0)(\nu_0-\nu_1)}  \\
\nu_1-\nu_0+\fr{\nu_0(\Gz_1-\Gz_0)}{s-\Gz_0} & -\Gz_1+\fr{\nu_0(\Gz_1-\Gz_0)s}{(\nu_0-\nu_1)(s-\Gz_0)}  \\
\end{array}\right).
\label{3.73}
\eeq 
The orders of the columns of the matrix $\tilde X(s)$ and its determinant at infinity are equal to
0. The determinant of the matrix $\tilde X(s)$ does not have zeros in any finite complex plane. Therefore
the matrix $\tilde X(s)$ is the canonical matrix of factorization 
and the partial indices $\Gk_1=\Gk_2=0$. 

Notice that the original vector Riemann-Hilbert problem can be rewritten as
\beq
\GF^+(t)=-[G(t)]^{-1}\GF^-(t)+[G(t)]^{-1}g(t), \quad t\in L.
\label{3.74}
\eeq
Since 
\beq
\tilde X(s)=[X(s)]^{-1}T(s), \quad s\in {\Bbb C},
\label{3.75}
\eeq
and 
\beq
G(t)=X^+(t)[X^-(t)]^{-1}=[X^-(t)]^{-1}X^+(t), \quad t\in L,
\label{3.76}
\eeq
we see that  
\beq
[G(t)]^{-1}=\tilde X^+(t)[T(t)]^{-1}T(t)[\tilde X^-(t)]^{-1}=\tilde X^+(t)[\tilde X^-(t)]^{-1}, \quad t\in L.
\label{3.77}
\eeq
Therefore $\tilde X(s)$ is the canonical matrix of factorization of the coefficient $[G(t)]^{-1}$ of the Riemann-Hilbert
problem (\ref{3.74}).
We may conclude now that the vector Riemann-Hilbert problem (\ref{3.12}) has zero partial indices and
according to the stability criterion  {\bf(\ref{goh})}, {\bf(\ref{vek})} they are stable.

\setcounter{equation}{0}
  
\section{Vector Riemann-Hilbert problem associated with the direct extension of the boundary conditions}\label{vec}

In the preceding section we solved the vector Riemann-Hilbert problem derived by employing the Laplace transforms 
of the velocity potentials.
In this section we wish to apply a different approach for its derivation that employs the Fourier transform
and another way of extension of the boundary conditions on the whole real axis.  We aim  to understand whether one method has the advantage of the other. 
For simplicity, we confine ourselves to the case $0<\Gt_0<\pi/2$ and take $L$ as the real axis. 
We start with writing the general integral representation of the  scattering potentials 
$$
\Gf_0(x,y)=\fr{1}{2\pi}\int_{-\infty}^\infty A_0(s) e^{-isx+\Gg y}ds, \quad y<0,
$$
\beq
\Gf_1(x,y)=\fr{1}{2\pi}\int_{-\infty}^\infty A_1(s) e^{-isx-\Gg y}ds, \quad y>0,
\label{4.1}
\eeq
where $\Gg=\Gg(s)$ is the branch fixed by (\ref{3.3}). The first derivative $\fr{\Md }{\Md y}\Gf_0(x,0)$ 
is continuous at the point $x=0$ and equals 0, while the mixed derivative $\fr{\Md^2 }{\Md x\Md y}\Gf_0(x,0)$ is bounded 
at the point $x=0$ and discontinuous,
\beq
\lim_{x\to 0^-} \fr{\Md^2 }{\Md x\Md y}\Gf_0(x,0)=0, \quad \lim_{x\to 0^+} \fr{\Md^2 }{\Md x\Md y}\Gf_0(x,0)=N,
\label{4.2}
\eeq
where $N$ is a nonzero constant. 
Extend now the boundary conditions (\ref{2.11}) onto the whole real axis $L=\{-\infty<x<\infty, y=0\}$ except for the point
$x=0$,
$$
\fr{\Md \Gf_0}{\Md y}=\Gf_0^+(x), \quad \fr{\Md \Gf_1}{\Md y}=\Gf_1^+(x), 
$$$$
\fr{\Md^3\Gf_0}{\Md x^2\Md y}+\mu^2\fr{\Md \Gf_0}{\Md y}-\Ga(\Gf_1-\Gf_0)=\Gb^+(x)+\Gf_0^-(x),
$$
\beq
\fr{\Md \Gf_1}{\Md y}-\fr{\Md \Gf_0}{\Md y}+k\tau\Gf_1=\Gf_1^-(x),  
\label{4.4}
\eeq
where 
$$
\Gf_j^+(x)=0, \quad x<0;  \quad \Gf_j^-(x)=0, \quad x>0,   \quad j=0,1.
$$
\beq
\Gb^+(x)=\left\{
 \begin{array}{cc}
-2\Ga e^{ikx\sin\Gt_0}, & x> 0,\\
0, & x<0.\\
\end{array}
\right.
\label{4.5}
\eeq
To derive the associated vector Riemann-Hilbert problem, we introduce the Laplace transforms (one-sided Fourier integrals)
\beq
\hat\GF_j^-(s)=\int_{-\infty}^0 \Gf_j^-(x)e^{i s x}dx, \quad \GF_j^+(s)=\int_0^{\infty}\Gf_j^+(x)e^{i s x}dx,\quad j=0,1,
\label{4.6}
\eeq
apply the Fourier integral transform to the boundary conditions (\ref{4.4}) and observe that
\beq
\int_{-\infty}^\infty\fr{\Md^3\Gf_0}{\Md x^2\Md y}(x,0)e^{isx}dx=-N-s^2\Gg(s)A_0(s).
\label{4.6'}
\eeq
Then the boundary condition (\ref{4.4}) can be rewritten in terms of the functions $\GF_j^\pm(s)$ and $A_j(s)$ as
$$
\Gg(s)A_0(s)=\GF_0^+(s), \quad \Gg(s)A_1(s)=-\GF_1^+(s), 
$$
$$
A_0(s)[\Gg(s)(\mu^2-s^2)+\Ga]-\Ga A_1(s)=\hat\GF_0^-(s)-\fr{2i\Ga}{s+k\sin\Gt_0}+N,
$$
\beq
[-\Gg(s)+k\tau]A_1(s)-\Gg(s)A_0(s)=\hat\GF_1^-(s).
\label{4.7}
\eeq
We now eliminate the functions $A_0(s)$ and $A_1(s)$ to have the vector Riemann-Hilbert problem
\beq
\left(\begin{array}{c}
\hat\GF_0^-(s)\\
\hat\GF_1^-(s)\\ 
\end{array}\right)=
\left(\begin{array}{cc}
\mu^2-s^2+ \fr{\Ga}{\Gg(s)} & \fr{\Ga}{\Gg(s)}\\
-1 & 1-\fr{k\tau}{\Gg(s)}\\
\end{array}\right)
\left(\begin{array}{c}
\GF_0^+(s)\\
\GF_1^+(s)\\ 
\end{array}\right)+\fr12 g(s),\quad s\in L,
\label{4.8}
\eeq
where
\beq
g(s)=\left(\fr{4i\Ga}{s+k\sin\Gt_0}-2N\right)\left(\begin{array}{c}
1\\
0\\ 
\end{array}\right).
\label{4.9}
\eeq
To transform the matrix coefficient of the problem to the form (\ref{3.10}), we replace the functions $\hat\GF^-_0(s)$
and  $\hat\GF^-_1(s)$ by two new functions,
\beq
\GF_0^-(s)=2\hat\GF_0^-(s)+\fr{2\Ga}{k\tau}\hat\GF_1^-(s), \quad
\GF_1^-(s)= -\fr{2\Ga}{k\tau}\hat\GF_1^-(s).
\label{4.10}
\eeq
In terms of the vector-functions $\GF^\pm(s)=(\GF_0^\pm(s),\GF_1^\pm(s))^\top$, the Riemann-Hilbert problem has the form 
\beq
\GF^-(s)=(\mu^2-s^2)G(s)\GF^+(s)+g(s), \quad s\in L.
\label{4.11}
\eeq
with the matrix coefficient $G(s)$ defined by
$$
G(s)=
\left(\begin{array}{cc}
b_0(s)+c_0(s)l_0(s), & c_0(s) m \\
c_0(s)m & b_0(s)-c_0(s)l_0(s)\\
\end{array}\right),
$$
\beq
b_0(s)=1+\fr{2\Ga}{\mu^2-s^2}\left(\fr{1}{\Gg}-\fr{1}{k\tau}\right),\quad 
c_0(s)=\fr{1}{\mu^2-s^2},
\quad 
l_0(s)=\mu^2-s^2, \quad m=\fr{2\Ga}{k\tau}.
\label{4.12}
\eeq
It is seen that the functions $b_0(s)$, $c_0(s)$ and $l_0(s)$ differ from the corresponding functions $b(s)$,  $c(s)$ and $l(s)$
appeared
in Section 3, while the characteristic polynomial $f(s)=(s^2-\mu^2)^2+m^2$ is the same. This means that both vector
Riemann-Hilbert problems reduce to the scalar Riemann-Hilbert problem (\ref{3.17}) on the same genus-1 Riemann surface $\CR$.
The solution of the problem (\ref{3.17}), the matrix factorization and Jacobi inversion problems are given by the same formulas as in Section 3 if
the functions $b(s)$, $c(s)$, and $l(s)$ are replaced by $b_0(s)$, $c_0(s)$, and $l_0(s)$, respectively.

Substantial differences between the two problems begin when we start analyzing the behavior of the functions 
$\GD(s)=\Gl_1(s)\Gl_2(s)$ and $\Ge(s)=\Gl_1(s)/\Gl_2(s)$
 at infinity. These asymptottics are required to determine the asymptotics of the the factorization matrix that is crucual
 for the application of the Liouville's  theorem.
It is not hard to show that
$$
\GD(t)=\fr{4\Ga}{k\tau t^2}\left(1-\fr{k\tau}{|t|}+O(t^{-2}\right),\quad t\to\pm \infty,
$$
\beq
\Ge(t)=\fr{\Ga}{k\tau t^2}\left(1-\fr{k\tau}{|t|}+O(t^{-2}\right),\quad t\to\pm \infty.
\label{4.13}
\eeq
This results in  logarithmic singularities of  the functions $\log\GD(t)$ and $\log\Ge(t)$ 
at infinity which make  the analysis of the behavior of the functions $\Gc_1(s)$ and $\Gc_2(s)$ at infinity harder.
Consider first the limit values  $\Gc^\pm_1(t)$ on the contour $L$ of the function  $\Gc_1(s)$ given by (\ref{3.51})
\beq
\Gc^\pm_1(t)=\pm\fr{\log\GD(t)}{4}+\fr{t}{2\pi i}
\int_0^1\fr{\log\GD(t_0) dt_0}{t^2_0-t^2}+
\fr12\int_\GG
\fr{dt_0}{t_0-t}+I_1(t)+I_2(t)+I_3(t),\quad t\in L,
\label{4.14}
\eeq
where
$$
I_1(t)=\fr{t}{2\pi i}\int_1^\infty\log\fr{k\tau t_0^2\GD(t_0)}{4\Ga}\fr{dt_0}{t^2_0-t^2},
$$
\beq
I_2(t)=\fr{t}{2\pi i}\log\fr{4\Ga}{k\tau}\int_1^\infty\fr{dt_0}{t^2_0-t^2}, \quad 
I_3(t)=-\fr{t}{\pi i}\int_1^\infty\fr{\log t_0 dt_0}{t_0^2-t^2}.
\label{4.15}
\eeq
Except for the last integral $I_3(t)$ the asymptotics at infinity of the terms in   (\ref{4.14})  and (\ref{4.15}) can be written immediately.
For the integrals $I_1(t)$ and $I_2(t)$ we have
$$
I_1(t)=\fr{k\tau}{2\pi i t}\log|t|-\fr{1}{2\pi it}\int_1^\infty\left(\fr{k\tau}{t_0}+\log\fr{k\tau t_0^2\GD(t_0)}{4\Ga}\right)dt_0+O(t^{-2}), \quad t\to\pm\infty,
$$
\beq
I_2(t)=\fr{1}{4\pi i}\log\fr{4\Ga}{k\tau}\log\fr{1+t^{-1}}{1-t^{-1}}=\fr{1}{2\pi i t}\log\fr{4\Ga}{k\tau}+O(t^{-3}), \quad t\to\pm\infty.
\label{4.15'}
\eeq
As for the integral $I_3(t)$, it can be expressed through the limit
\beq
I_3(t)=-\fr{t}{\pi i}\lim_{\nu\to 0}\fr{\CI_\nu(t)-\CI_0(t)}{\nu},\quad \CI_\nu(t)=\int_1^\infty\fr{t_0^\nu dt_0}{t_0^2-t^2}.
\label{4.16}
\eeq
On making the substitutions $t_0=y^{-1/2}$ and $|t|=x^{-1/2}$   we represent the integral  $ \CI_\nu(t)$ as a Mellin convolution integral
\beq
\CI_\nu (t)=-\fr{x}{2}\int_0^\infty \fr{h_-(y)}{1-\fr{x}{y}}\fr{dy}{y},\quad x>0,
\label{4.17}
\eeq
where
\beq
h_-(y)=\left\{\begin{array}{cc}
y^{-\nu/2-1/2}, & 0<y<1,\\
0, & y>1.\\
\end{array}
\right.
\label{4.18}
\eeq
By the Mellin convolution theorem we recast the integral to write
\beq
\CI_\nu (t)=-\fr{x}{4 i}\int_{\Gk-i\infty}^{\Gk+i\infty}\fr{\cot\pi \Gs}{\Gs-\fr{\nu+1}{2}} x^{-\Gs}d\Gs, \quad 
\max\left\{0,\fr{\nu+1}{2}\right\}<\Gk<1,
\label{4.19}
\eeq
and compute it by the theory of residues. In the case  $|t|>1$ we have
\beq
\CI_\nu(t)=\sum_{j=0}^\infty\fr{t^{-2j-2}}{2j+\nu+1}+\fr{\pi|t|^{\nu-1}}{2}\tan\fr{\pi \nu}{2}.
\label{4.20}
\eeq
If we substitute this expression into (\ref{4.17}) and compute the limit, we obtain
\beq
I_3(t)=\fr{\pi i}{4}\sgn t+\fr{1}{\pi i}\sum_{j=0}^\infty\fr{t^{-2j-1}}{(2j+1)^2}.
\label{4.21}
\eeq
When we combine the asymptotics we derived with those of the other terms in  (\ref{4.14}) and (\ref{4.15}) we get
the following representation of the functions $\Gc^\pm_1(t)$ on the real axis for large $|t|$:
\beq
\Gc_1^\pm(t)=\mp\fr12\left[\log|t|-\fr12\log\fr{4\Ga}{k\tau}+\fr{k\tau}{2|t|}\right]+\fr{\pi i}{4}\sgn t+\fr{kt}{2\pi i}\fr{\log|t|}{t}+\fr{r_1}{t}+O(t^{-2}),\quad 
t\to\pm\infty,
\label{4.22}
\eeq
where
\beq
r_1=\fr{1}{2\pi i}\left[
2+\log\fr{4\Ga}{k\tau}-\int_0^1\log\GD(t)dt
-\int_1^\infty\left(\fr{k\tau}{t}+
\log\fr{k\tau t^2\GD(t)}{4\Ga}\right)dt
\right]+\fr{\Gz_0-\Gz_1}{2}.
\label{4.23}
\eeq
Analyze now the behavior of the function $\Gc_1(s)$ as $s\to\infty$ and $s\in{\Bbb C^\pm}\setminus L$. 
We have
\beq
\Gc_1(s)=\fr{s}{2\pi i}
\int_0^1\fr{\log\GD(t) dt}{t^2-s^2}+
\fr12\int_\GG
\fr{dt}{t-s}+I_1(s)+I_2(s)+I_3(s),\quad s\in{\Bbb C^\pm}\setminus L,
\label{4.24}
\eeq
where $I_j(s)$ are defined by (\ref{4.15}). As before, we focus  our attention on the integral $I_3(s)$. On making the substitution
$s=\pm is_0$, $-\fr12\pi<\arg s_0<\fr12 \pi$, $s\in{\Bbb C^\pm}$ we represent the  integral $I_3(s)$ in the form
\beq
I_3(s)=-\fr{s}{\pi i}\lim_{\nu\to 0}\fr{\CJ_\nu(s_0)-\CJ_0(s_0)}{\nu},\quad \CJ_\nu(s_0)=\int_1^\infty\fr{t^\nu dt}{t^2+s_0^2}.
\label{4.25}
\eeq
Substitute next $t$ by $y^{-1/2}$ and $s_0^2$ by $x^{-1}$ and  write the integral $\CJ_\nu(s_0)$ as a Mellin convolution integral 
\beq
\CJ_\nu(s_0)=\fr{x}{2}\int_0^\infty \fr{h_-(y)}{1+\fr{x}{y}}\fr{dy}{y}.
 \label{4.26}
 \eeq 
We apply the convolution theorem and convert this integral into the following one:
\beq
\CJ_\nu (s_0)=\fr{x}{4 i}\int_{\Gk-i\infty}^{\Gk+i\infty}\fr{x^{-\Gs}}{\sin\pi \Gs}
\fr{d\Gs}{\Gs-\fr{\nu+1}{2}},  \quad 
\max\left\{0,\fr{\nu+1}{2}\right\}<\Gk<1.
\label{4.27}
\eeq
After the theory of residues is employed we have a series representation of the integral for $|s_0|>1$
\beq
\CI_\nu(s_0)=\sum_{j=0}^\infty\fr{(-1)^{j+1}x^{j+1}}{2j+\nu+1}+\fr{\pi x^{1/2-\nu/2}}{2\cos\fr{\pi \nu}{2}}, \quad x=s_0^{-2}, \quad |\arg s_0|<\fr{\pi}{2}.
\label{4.28}
\eeq
On coming back to $s$ ($s_0 =\mp is$, $s\in{\Bbb C^\pm}$) and computing the limit in (\ref{4.25}) we obtain
\beq
I_3(s)=\mp\fr12\log(\mp is)+\fr{1}{\pi i}\sum_{j=0}^\infty\fr{s^{-2j-1}}{(2j+1)^2}, \quad |s|>1, \quad s\in{\Bbb C^\pm}.
\label{4.29}
\eeq
The asymptotics of the other terms in (\ref{4.24}) is derived in a simple manner, and we have
\beq
\Gc_1(s)=\mp\fr12\log(\mp is)\pm\fr14\log\fr{4\Ga}{k\tau}+\fr{k\tau}{2\pi i s}\log(\mp is)+\fr{r_1}{s}+O(s^{-2}),\quad 
s\to\infty,\quad s\in {\Bbb C^\pm},
\label{4.30}
\eeq
where $r_1$ is given by (\ref{4.23}).
It becomes evident that when $s\to t\in L$ in (\ref{4.30}), then for both cases $t>0$ and $t<0$ treated separately, the asymtpotics
deduced coincide with formula (\ref{4.22}).

The derivation of the asymptotics of the function $f^{1/2}(s)\Gc_2(s)$ as $s\to\infty$ is analogous to the deduction of the asymptotics
of the function $\Gc_1(s)$. We have
$$
f^{1/2}(s)\Gc_2(s)=\mp\fr12\log(\mp s)+\fr{k\tau}{2\pi i s}\log(\mp is)\pm \fr{1}{4}\log\fr{\Ga}{k\tau}
$$
\beq
-\fr12\left(\int_\GG+m_a\int_{\Ba}+m_b\int_\Bb\right)\fr{tdt}{\Gx(t)}+\fr{r_2}{s}+O(s^{-2}),\quad 
s\to\infty, \quad s\in{\Bbb C^\pm},
\label{4.31}
\eeq
where
$$
r_2=\fr{1}{2\pi i}\left[
2+\log\fr{\Ga}{k\tau}-\int_0^1\log\Ge(t)\fr{t^2dt}{f^{1/2}(t)}
-\int_1^\infty\left(\fr{k\tau}{t}+
\fr{t^2}{f^{1/2}(t)}\log\fr{k\tau t^2\Ge(t)}{\Ga}\right)dt
\right]
$$
\beq
-\fr12\left(\int_\GG+m_a\int_{\Ba}+m_b\int_\Bb\right)\fr{t^2dt}{\Gx(t)}.
\label{4.32}
\eeq
As before for the function $\Gc_1(s)$, the two asymptotics for large $t$ when $t\in L$ derived by means of the Sokhotski-Plemelj formulas and directly
from the asymptotics (\ref{4.31}) as $s\to t\pm 0$ coincide. 

Our next step is to write down the asymptotics of the solution 
\beq
F(s,\pm f^{1/2}(s))=e^{\Gc_1(s)\pm f^{1/2}(s)\Gc_2(s)}
\label{4.33}
\eeq
of the scalar Riemann-Hilbert problem
on the surface $\CR$. We have
$$
F(s, f^{1/2}(s))=e^{r_1^\pm+r_2^\pm}(\mp is)^{\mp 1}
\left[1+\fr{k\tau}{\pi i s}\log(\mp is)+\fr{r_1+r_2}{s} + O\left(\fr{\log^2 s}{s^2}\right)\right],
$$
\beq
F(s,-f^{1/2}(s))=
e^{r_1^\pm-r_2^\pm}
\left[
1+\fr{r_1-r_2}{s}+
O\left(\fr{1}{s^2}\right)
\right], \quad s\to\infty, \quad s\in{\Bbb C^\pm},
\label{4.34}
\eeq
where
\beq
r_1^\pm=\pm \fr{1}{4}\log\fr{4\Ga}{k\tau},
\quad 
r_2^\pm=\pm \fr{1}{4}\log\fr{\Ga}{k\tau}-\fr12\left(\int_\GG+m_a\int_{\Ba}+m_b\int_\Bb\right)\fr{tdt}{\Gx(t)}.
\label{4.35}
\eeq
On substituting the asymptotics (\ref{4.34}) into the expressions (\ref{3.15}), (\ref{3.16}), where $l(s)$ needs to be replaced by 
$l_0(s)$, we deduce
$$
X^\pm(s)=e^+_\pm(\mp is)^{\mp 1}\left[1+\fr{k\tau}{\pi is}\log(\mp s)+\fr{\Gk_+}{s}+O\left(\fr{\log^2 s}{s^2}\right)\right]
\left[\left(\begin{array}{cc}
0 & 0\\
0 & 1
\end{array}\right)+O(s^{-2})\right]
$$$$
+e^-_\pm\left(1+\fr{\Gk_-}{s}\right)
\left[\left(\begin{array}{cc}
1 & 0\\
0 & 0
\end{array}\right)+O(s^{-2})\right], \; s\to\infty, \; s\in {\Bbb C^\pm},
$$
$$
[X^\pm(s)]^{-1}=\fr{1}{e^+_\pm}(\mp is)^{\pm 1}\left[1-\fr{k\tau}{\pi is}\log(\mp s)-\fr{\Gk_+}{s}+O\left(\fr{\log^2 s}{s^2}\right)\right]
\left[\left(\begin{array}{cc}
0 & 0\\
0 & 1
\end{array}\right)+O(s^{-2})\right]
$$
\beq
+\fr{1}{e^-_\pm}\left(1-\fr{\Gk_-}{s}\right)
\left[\left(\begin{array}{cc}
1 & 0\\
0 & 0
\end{array}\right)+O(s^{-2})\right], \; s\to\infty, \; s\in {\Bbb C^\pm},
\label{4.36}
\eeq
where
\beq
e_\pm^+=e^{r_1^\pm+r_2^\pm}, \quad e_\pm^-=e^{r_1^\pm-r_2^\pm}, \quad 
\Gk_\pm =r_1\pm r_2.
\label{4.37}
\eeq
Now we replace $G(s)$ by $[X^-(s)]^{-1}X^+(s)$, $s\in L$, in (\ref{4.11}) and represent $X^-(s)g(s) $ as 
\beq
X^-(s)g(s)=\Psi^+(s)-\Psi^-(s), 
\label{4.38}
\eeq
where
$$
\Psi^+(s)=\fr{4\Ga i}{(s+k\sin\Gt_0)(\mu+k\sin\Gt_0)}
\left(\begin{array}{c}
X^-_{1}(-k\sin\Gt_0)\\
X^-_{2}(-k\sin\Gt_0)\\
\end{array}\right),
$$
\beq
\Psi^-(s)=\fr{4\Ga i}{s+k\sin\Gt_0}\left[\fr{1}{\mu+k\sin\Gt_0}
\left(\begin{array}{c}
X^-_{1}(-k\sin\Gt_0)\\
X^-_{2}(-k\sin\Gt_0)\\
\end{array}\right)-\fr{1}{\mu-s}
\left(\begin{array}{c}
X^-_{1}(s)\\
X^-_{2}(s)\\
\end{array}\right)\right],
\label{4.39}
\eeq
and 
$$
X_1^-(s)=e^{\Gc_1^-(s)}[\cosh(f^{1/2}(s)\Gc_2^-(s))+\fr{\mu^2-s^2}{f^{1/2}(s)}\sinh(f^{1/2}(s)\Gc_2^-(s))],
$$
\beq
X_2^-(s)=\fr{me^{\Gc_1^-(s)}}{f^{1/2}(s)}\sinh(f^{1/2}(s)\Gc_2^-(s)).
\label{4.40}
\eeq
Return next to the boundary condition of the Riemann-Hilbert problem. We have
\beq
\fr{1}{\mu-s}X^-(s)\left[\GF^-(s)+2N
\left(\begin{array}{c}
1\\
0\\
\end{array}\right)\right]+\GY^-(s)=(\mu+s)X^+(s)\GF^+(s)+\GY^+(s), \quad s\in L.
\label{4.41}
\eeq
Using the asymptotics (\ref{4.36}) of the matrices $X^\pm(s)$ at infinity we conclude
from  (\ref{4.41}) that 
$$
\GF_0^-(s)=O(s^{-1}), \quad \GF_1^-(s)=O(s^{-1}), \quad s\to\infty, \quad s\in {\Bbb C^-}, 
$$
\beq
\GF_0^+(s)=O(s^{-2}), \quad \GF_1^+(s)=O(s^{-1}), \quad s\to\infty, \quad s\in {\Bbb C^+}.
\label{4.42}
\eeq
The asymptotics of $\GF^+(s)=O(s^{-2})$, $s\to\infty$, $s\in{\Bbb C^+}$ results in $\Gf^+(x)\to 0$, $x\to 0^+$,
and the continuity condition for the displacement at $x=0$, $y=0^-$ is fulfilled. 

As in Section 3, the vector $X(s)\GF(s)$ has a simple pole at the point $\Gz_0$ and admits the representation (\ref{3.43}).
By applying the continuity principle and the generalized Liouville theorem we find the solution
of the vector Riemann-Hilbert problem
$$
\GF^-(s)= -2N
\left(\begin{array}{c}
1\\
0\\
\end{array}\right)+(\mu-s)[X^-(s)]^{-1}\left[
\fr{C}{s-\Gz_0}\left(\begin{array}{c}
1\\
\Gn_0\\
\end{array}\right)-\GY^-(s)\right], \quad s\in{\Bbb C^-},
$$
\beq
\GF^+(s)= \fr{1}{\mu+s}[X^+(s)]^{-1}\left[
\fr{C}{s-\Gz_0}\left(\begin{array}{c}
1\\
\Gn_0\\
\end{array}\right)-\GY^+(s)\right], \quad s\in{\Bbb C^+},
\label{4.43}
\eeq
where $C$ is an arbitrary constant. 
Now, the solution we derived has an unacceptable simple pole at the point $(\Gz_1,w_1)\in\CR$. It becomes a
removable point if the constant $C$ is fixed by the condition
\beq
C=\fr{\Gz_1-\Gz_0}{w_1+l(\Gz_1)+m\Gn_0}\left[(w_1+l(\Gz_1))\Psi_0(\Gz_1)+m\Psi_1(\Gz_1)\right].
\label{4.44}
\eeq
Finally, we verify the asymptotics of the solution (\ref{4.43}) at infinity. On substituting 
 the representation of the matrix $[X^-(s)]^{-1}$ from   rd(\ref{4.36}) into formulas (\ref{4.43}) we determine
 $$
 \GF_0^-(s)=-2N-\fr{C-\hat\GY_0}{e_-^-}+O(s^{-1}), \quad \GF_1^-(s)\sim\fr{i}{s e_-^+}(C\Gn_0-\hat\GY_1), \quad s\to\infty, \quad s\in{\Bbb C^-},
$$
\beq
 \GF_0^+(s)\sim\fr{i}{s^2 e_+^-}(C-\hat\GY_0), \quad  \GF_1^+(s)\sim-\fr{i}{s e_+^+}(C\Gn_0-\hat\GY_1), 
 \quad s\to\infty, \quad s\in{\Bbb C^+},
 \label{4.45}
 \eeq
where
\beq
\hat\GY_j=\fr{4\Ga i}{\mu+k\sin\Gt_0}X_j^-(-k\sin\Gt_0), \quad j=0,1.
\label{4.46}
\eeq
It is clear that the function $\GF_0^-(s)$ vanishes at infinity,  $\GF_0^-(s)=O(s^{-1})$, $s\to\infty$, $s\in{\Bbb C^-}$, if and only if
the constant $N=\fr{\Md^2 }{\Md x\Md y}\Gf_0(0^+,0)$ is fixed as
\beq
N=-\fr{C-\hat\GY_0}{2e_-^-}.
\label{4.47}
\eeq
This completes the solution of the vector Riemann-Hilbert problem (\ref{4.11}), (\ref{4.12}).

\setcounter{equation}{0}

\section{Numerical results}

\begin{figure}[t]
\centerline{
\scalebox{0.5}{\includegraphics{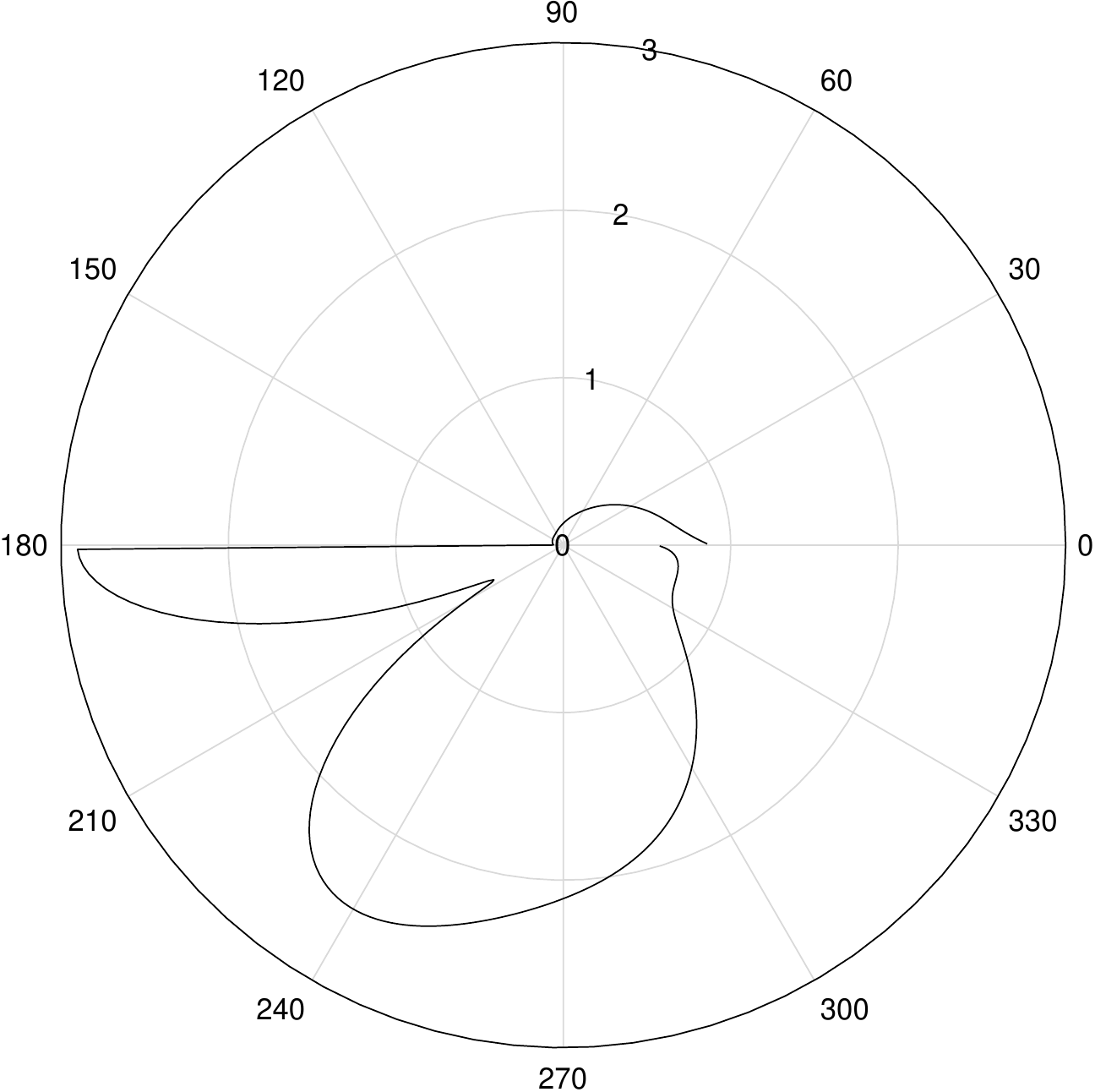}}}
\caption{
Variation of the function $P(r,\Gt)$ for $r=5$ and $0\le \Gt\le 2\pi$ when $\theta_0=\pi/4$, 
$\Gr_f/m_0=100$,  $d=d_1=d_2=0.01$, $a=0.001$, $|k|=1$, $\arg k=\tan^{-1} 0.1$, $|\Ga|=10$.}
\label{fig3}
\end{figure}

\begin{figure}[t]
\centerline{
\scalebox{0.5}{\includegraphics{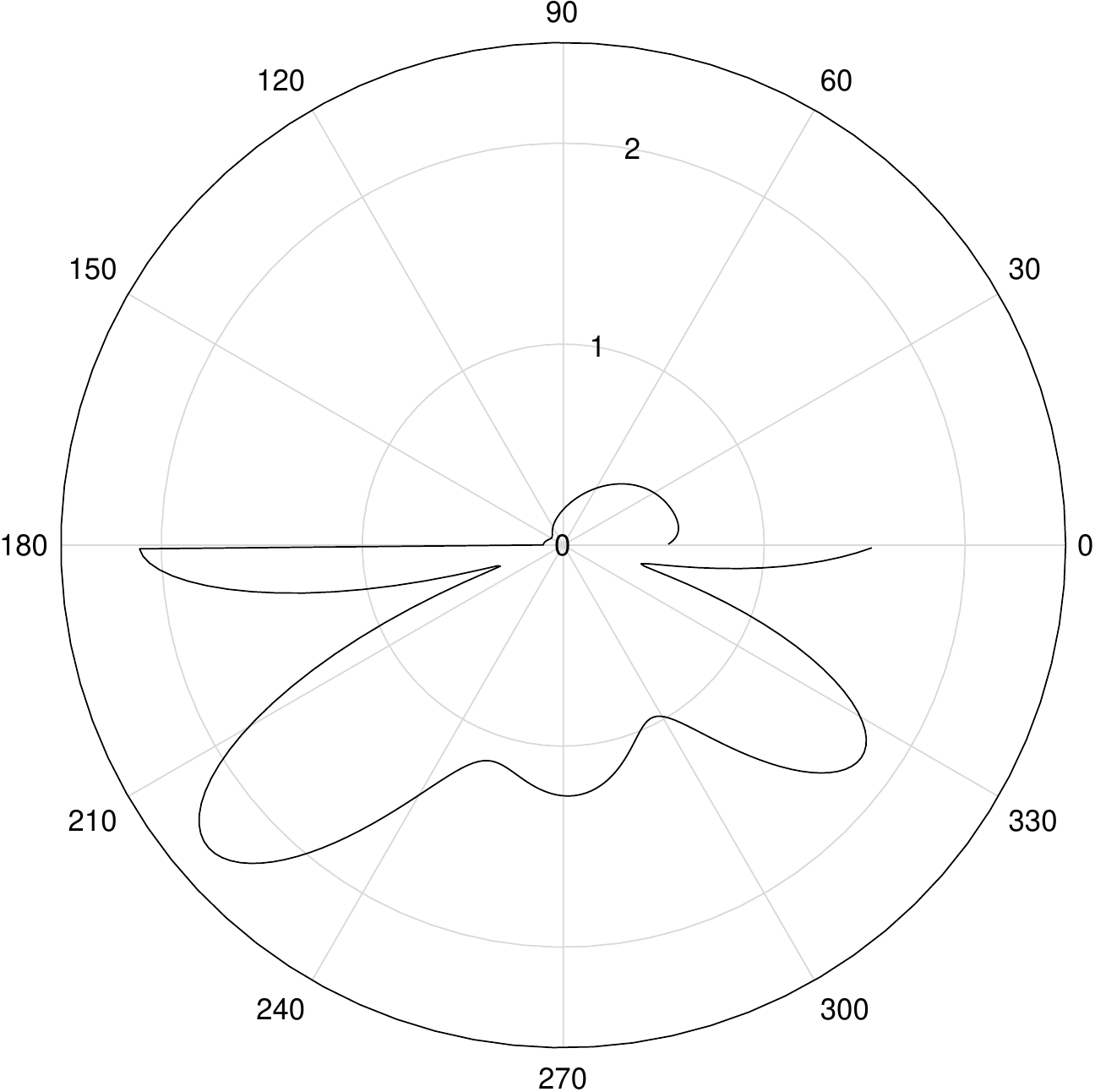}}}
\caption{Variation of the function $P(r,\Gt)$ for $r=5$ and $0\le \Gt\le 2\pi$ when $\theta_0=\pi/16$, 
$\Gr_f/m_0=100$,  $d=d_1=d_2=0.01$, $a=0.001$, $|k|=1$, $\arg k=\tan^{-1} 0.1$, $|\Ga|=10$.}
\label{fig4}
\end{figure}

\begin{figure}[t]
\centerline{
\scalebox{0.5}{\includegraphics{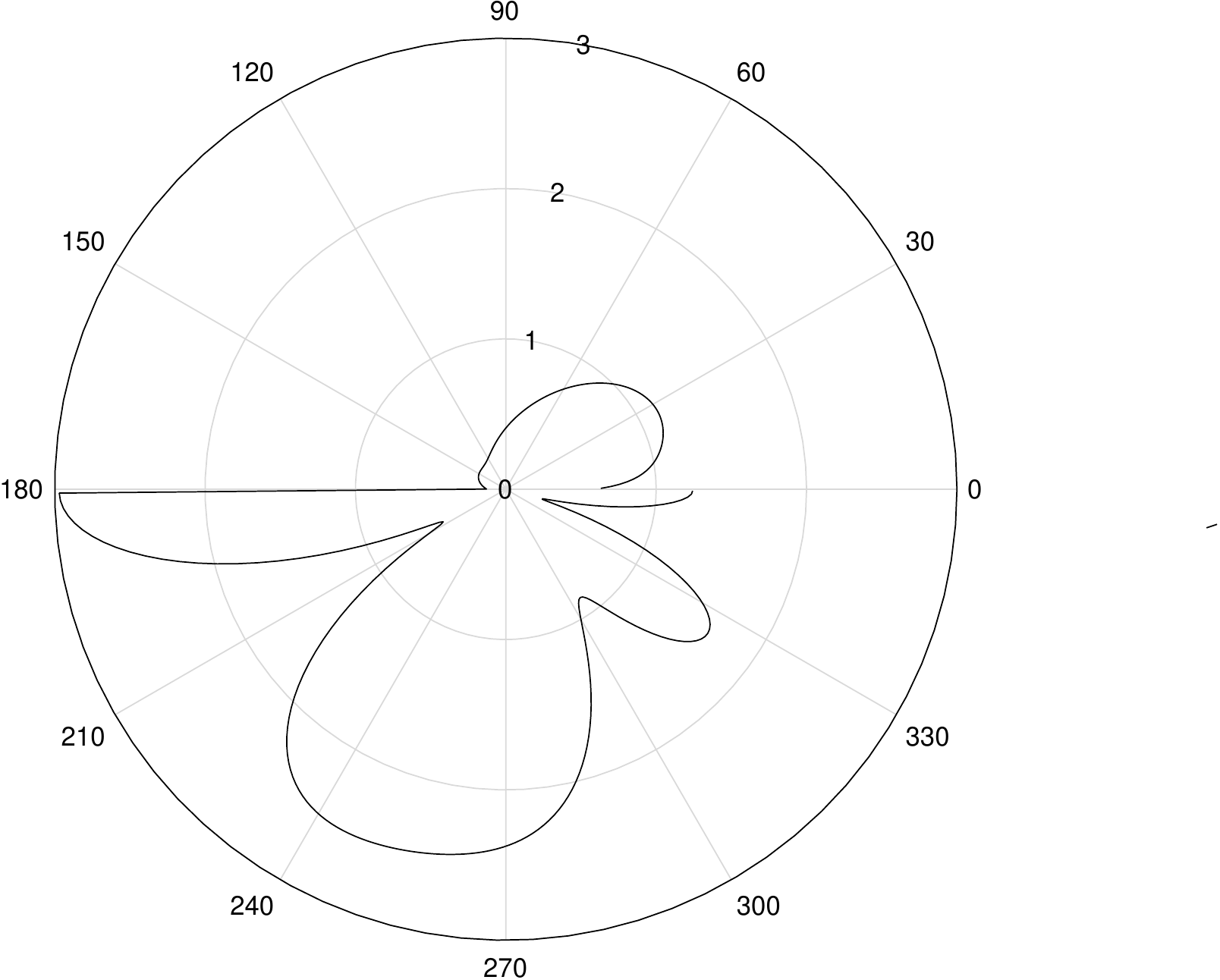}}}
\caption{Variation of the function $P(r,\Gt)$ for $r=5$ and $0\le \Gt\le 2\pi$ when $\Gr_f/m_0=500$, 
$\theta_0=\pi/4$, $|\Ga|=10$,  $d=d_1=d_2=0.01$, $a=0.001$, $|k|=1$, $\arg k=\tan^{-1} 0.1$, $|\Ga|=10$.}
\label{fig5}
\end{figure} 

\begin{figure}[t]
\centerline{
\scalebox{0.5}{\includegraphics{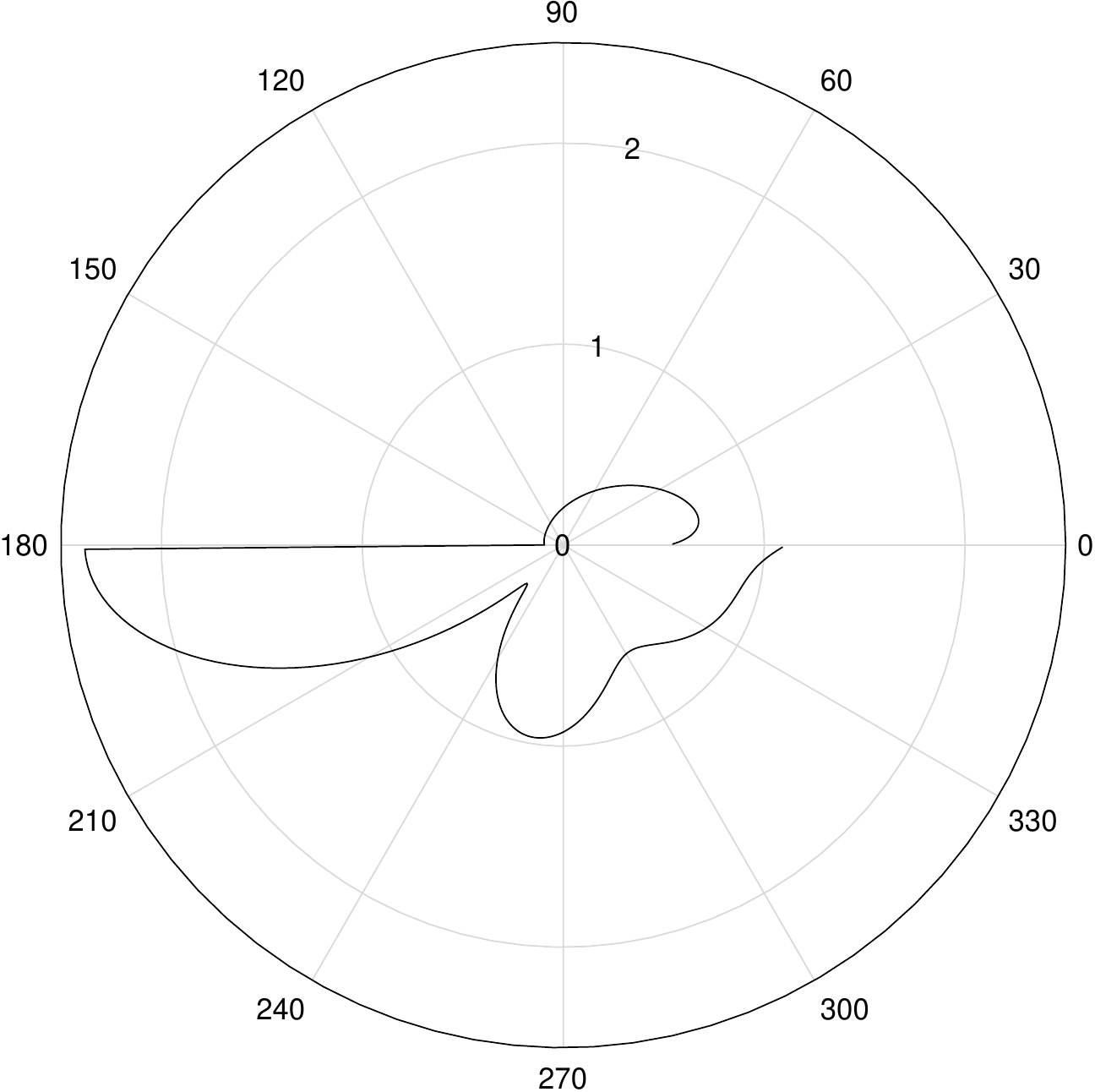}}}
\caption{Variation of the function $P(r,\Gt)$ for $r=3$ and $0\le \Gt\le 2\pi$ when $\theta_0=\pi/4$, 
$\Gr_f/m_0=100$,  $d=d_1=d_2=0.01$, $a=0.001$, $|k|=1$, $\arg k=\tan^{-1} 0.1$, $|\Ga|=10$.}
\label{fig6}
\end{figure}

\begin{figure}[t]
\centerline{
\scalebox{0.5}{\includegraphics{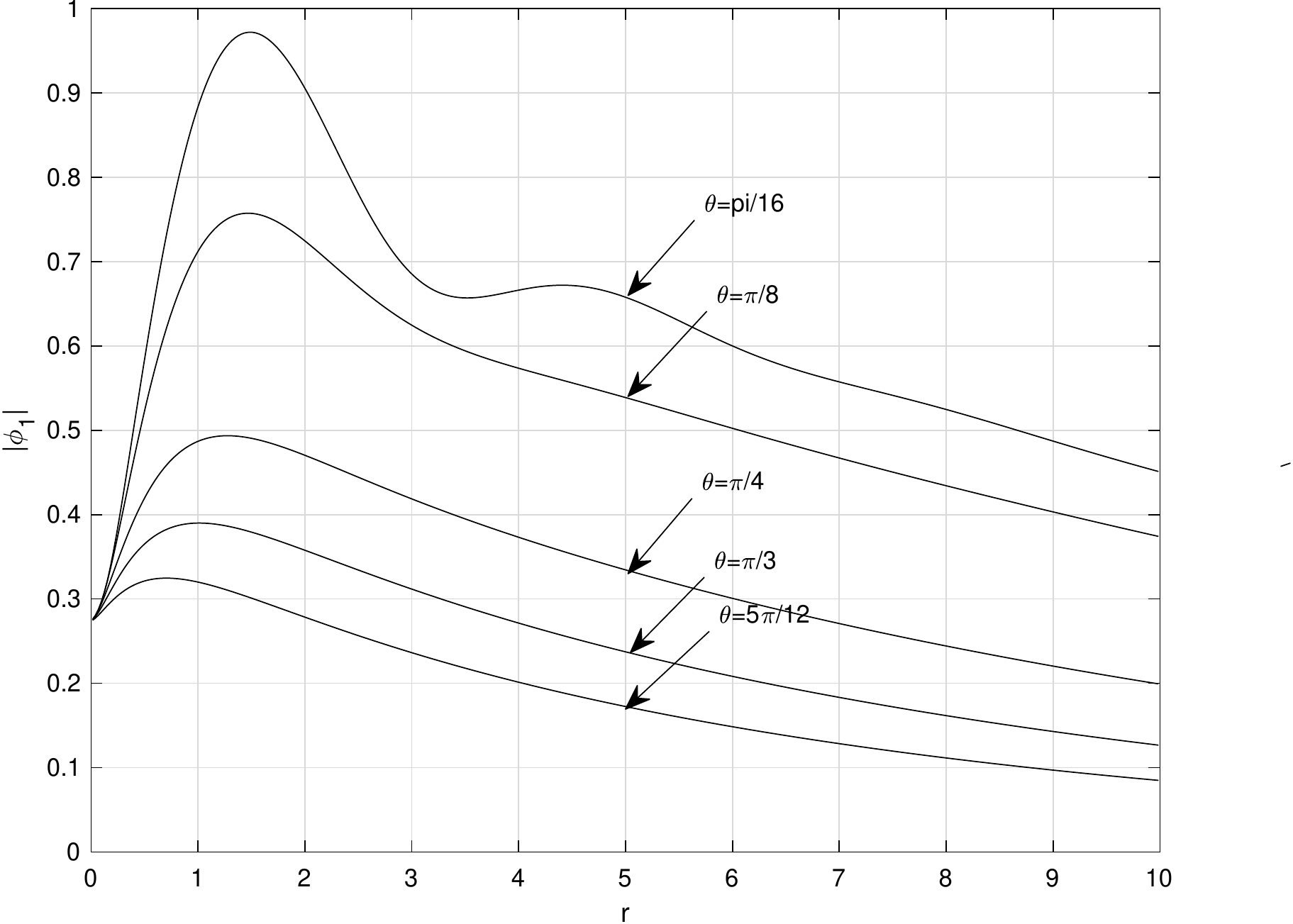}}}
\caption{
Variation of the function $P(r,\Gt)=|\Gf_1(x,y)|$ versus  $r$ for $\Gt=\pi/16, \pi/8, \pi/4, \pi/3$,  and $5\pi/12$ when 
$\theta_0=\pi/4$, 
$\Gr_f/m_0=100$,  $d=d_1=d_2=0.01$, $a=0.001$, $|k|=1$, $\arg k=\tan^{-1} 0.1$, $|\Ga|=10$.}
\label{fig7}
\end{figure}

\begin{figure}[t]
\centerline{
\scalebox{0.5}{\includegraphics{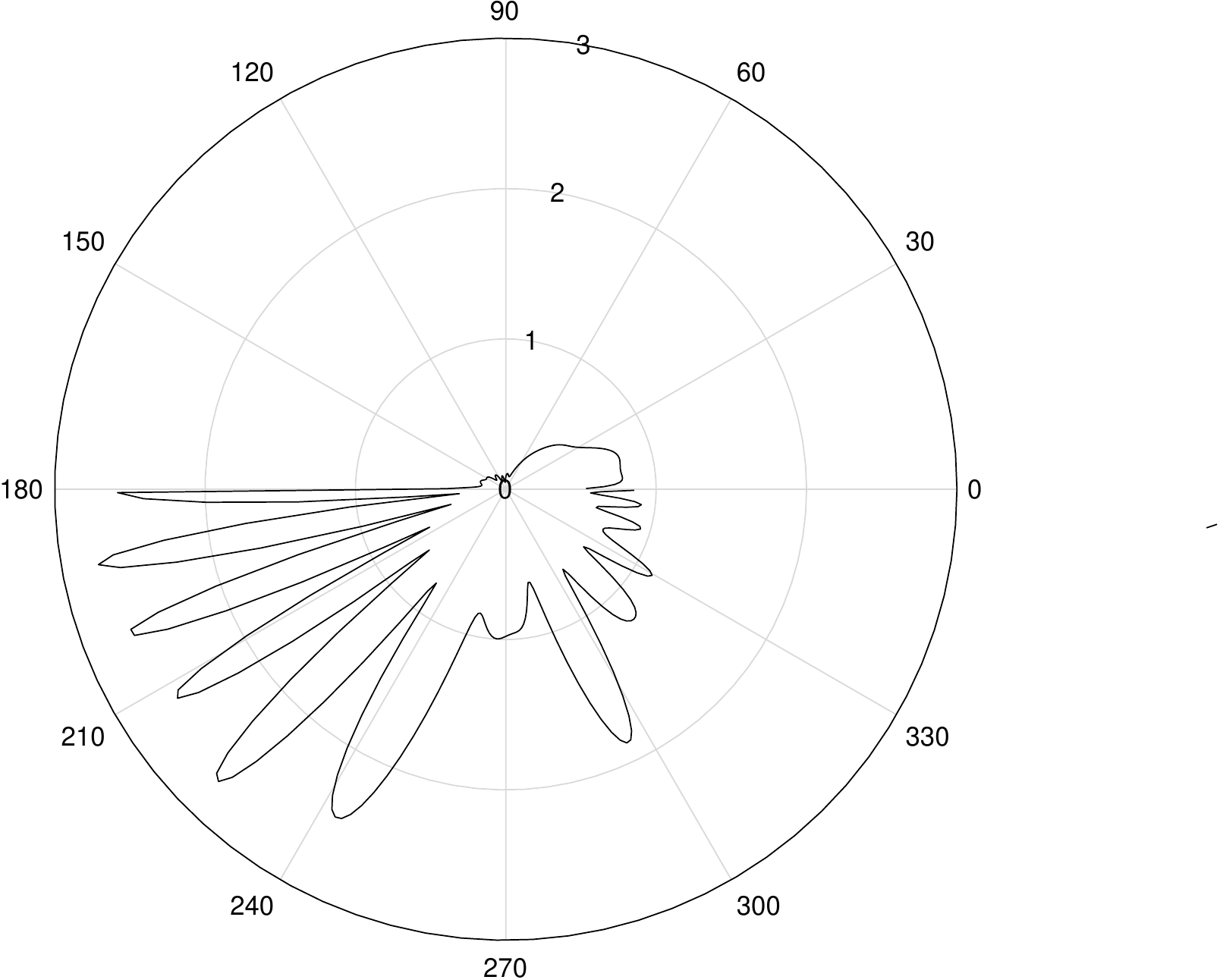}}}
\caption{
Variation of the function $P(r,\Gt)$ for 
 $r=5$ and $0\le \Gt\le 2\pi$ when $|k|=5$, $\arg k=\tan^{-1} 0.02$, $|\Ga|=250$,
  $\theta_0=\pi/4$, 
$\Gr_f/m_0=100$,  $d=d_1=d_2=0.01$, $a=0.001$.}
\label{fig8}
\end{figure}

\begin{figure}[t]
\centerline{
\scalebox{0.5}{\includegraphics{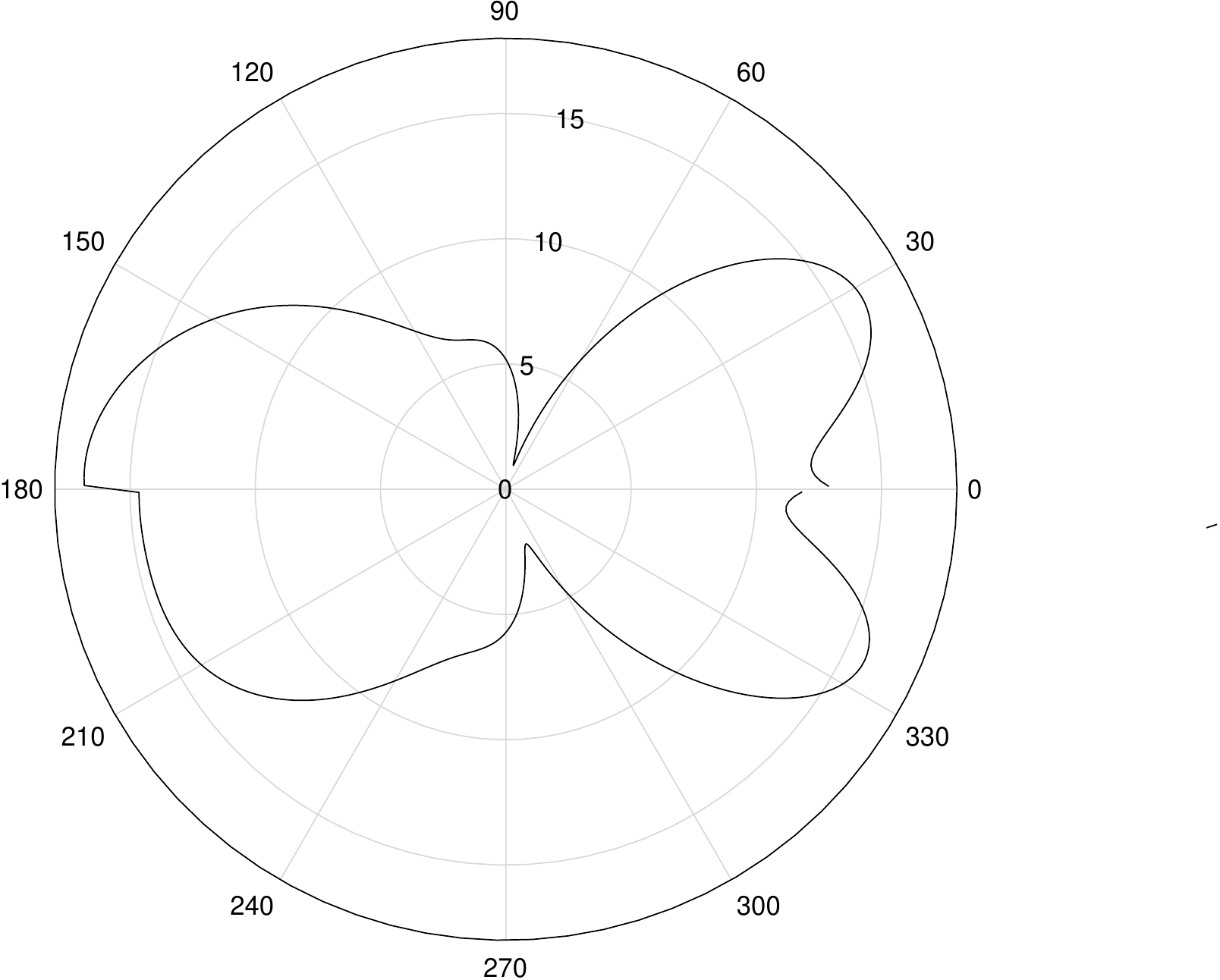}}}
\caption{
Variation of the function $P(r,\Gt)$ when $r=5$, $0\le \Gt\le 2\pi$ and $k$ is close to the resonance value $k_{res}$. 
The parameters are $\theta_0=\pi/4$, 
$\Gr_f/m_0=100$,  $d=d_1=d_2=0.2$, $a=0.005$, $|k|=1$, $\arg k=\tan^{-1} 0.02$, $|\Ga|=10.$}
\label{fig9}
\end{figure}

For our computations, we shall use the solution obtained in Section 3 and consider the case  $0<\Gt_0<\pi/2$. 
In this case we may choose $\Gk_0=0$, and the contour $L$ is the real axis.  By inverting the integrals (\ref{3.4}) and
employing formula (\ref{3.7}) we express 
the two potential $\Gf_0(x,y)$ and $\Gf_1(x,y)$ through the solution of the Riemann-Hilbert problem 
(\ref{3.12}) to (\ref{3.13.0})
$$
\Gf_0(x,y)=\fr{1}{2\pi}\int_{-\infty}^\infty[\GF_{0+}(s,0)+\GF_{0-}(s,0)]e^{\Gg y-is x}ds, \quad  y<0.
$$
\beq
\Gf_1(x,y)=\fr{1}{2\pi}\int_{-\infty}^\infty[\GF_{1+}(s,0)+\GF_{1-}(s,0)]e^{-\Gg y-is x}ds, \quad  y>0.
\label{5.1}
\eeq
On substituting formulas  (\ref{3.46'}) and (\ref{3.57})  into these expressions we transform them as
$$
\Gf_0(x,y)=C\hat\GO_0(r,\Gt)+N\hat\GO_1(r,\Gt)+\hat\GO_2(r,\Gt), \quad  y<0,
$$
\beq
\Gf_1(x,y)=C\tilde\GO_0(r,\Gt)+N\tilde\GO_1(r,\Gt)+\tilde\GO_2(r,\Gt), \quad  y>0,
\label{5.2}
\eeq
where $(r,\Gt)$ are the polar coordinates of a point $(x,y)$, $x=r\cos\Gt$, $y=r\sin\Gt$, 
$$
\hat\GO_0(r,\Gt)=\fr{1}{2\pi}\int_{-\infty}^\infty[\GL_1^+(t)-\GL_1^-(t)-\Gn_0\GL_2^+(t)+\Gn_0\GL_2^-(t)]\fr{v_+(t,r,\Gt)dt}{t-\Gz_0},
$$
$$
\hat\GO_j(r,\Gt)=\fr{1}{2\pi}\int_{-\infty}^\infty[\GL_1^+(t)\GY_{0+}^{(j)}(t)-\GL_{1}^-(t)\GY_{0-}^{(j)}(t)-\GL_2^+(t)\GY_{1+}^{(j)}(t)+\GL_2^-(t)\GY_{1-}^{(j)}(t)]v_+(t,r,\Gt)dt,\
$$
$$
\tilde\GO_0(r,\Gt)=\fr{1}{2\pi}\int_{-\infty}^\infty[\GL_2^-(t)-\GL_2^+(t)-\Gn_0\GL_0^-(t)+\Gn_0\GL_0^+(t)]\fr{v_-(t,r,\Gt)dt}{t-\Gz_0},
$$
$$
\tilde\GO_j(r,\Gt)=\fr{1}{2\pi}\int_{-\infty}^\infty[\GL_2^-(t)\GY_{0-}^{(j)}(t)-\GL_{2}^+(t)\GY_{0+}^{(j)}(t)-\GL_0^-(t)\GY_{1-}^{(j)}(t)+\GL_0^+(t)\GY_{1+}^{(j)}(t)]v_-(t,r,\Gt)dt,
$$
\beq
 j=1,2, \quad v_\pm(t,r,\Gt)=e^{[\pm\Gg(t)\sin\Gt-it\cos\Gt]r},
\label{5.3}
\eeq
and the functions $\GL_j^\pm(t)$ are given by (\ref{3.58}).

The improper integrals over the real axis $L$ are computed by mapping the contour $L$ onto the unit, positively oriented circle 
$\CC=\{|u|=1\}$ and applying the Simpson rule,
\beq
\int_L S(t)dt=-4\int_\CC \fr{S(t)d u}{(u-i)^2}=-4i\int_{\pi/2}^{5\pi/2}\fr{S(t) ud\Gt}{(u-i)^2},
\label{5.4}
\eeq
where $t=-2i(u+i)(u-i)^{-1}$ and $u=e^{i\Gt}$.
The principal value of the Cauchy integrals over the contour $L$ are 
computed by employing the same mapping to the unit circle $\CC$ and the following formula:
$$
\fr{1}{2\pi i}\int_L\fr{S(t)dt}{t-s}=\fr{\Gs-i}{2\pi i}\int_\CC\fr{S(t)du}{(u-i)(u-\Gs)}
$$
\beq
=\fr{\Gs-i}{2(2n+1)}\sum_{j=-n}^n\fr{S(t_j)}{u_j-i}\left[1+\fr{2i\sin((\Gt-\Gt_j)n/2)\sin((\Gt-\Gt_j)(n+1)/2)}{\sin((\Gt-\Gt_j)/2)}\right],
\quad s\in L,
\label{5.5}
\eeq
where 
\beq
\Gs=i\fr{s-2i}{s+2i}, \quad \Gt=-i\log\Gs, \quad t_j=-2i\fr{u_j+i}{u_j-i}, \quad 
u_j=e^{i\Gt_j}, \quad \Gt_j=\fr{2\pi j}{2n+1},
\label{5.6}
\eeq
and $n$ is the number of knots of the integration formula. For computing integrals (\ref{3.25}), (\ref{3.34'}), and
(\ref{3.36}) we apply the Gauss quadrature formula with Chebyshev's weights and abscissas.

In our numerical tests, we focus our attention on the absolute values of the full potentials 
$\Gy_0=\Gy_{inc}+\Gf_{ref}+\Gf_0$ in the lower half-plane ${\Bbb H_0}$ and  $\Gy_1=\Gf_1$ in the upper half-plane ${\Bbb H_1}$.
Denote by $P(r,\Gt)=|\Gy_1(x,y)|$, $(x,y)\in {\Bbb H_1}$ ($0<\Gt<\pi$) and 
$P(r,\Gt)=|\Gy_0(x,y)|$, $(x,y)\in {\Bbb H_0}$ ($\pi<\Gt<2\pi$).
For all tests we choose water's density $\Gr_f=997$ kg/m$^3$.   Except for  Figs. 8 and 9, we choose $\arg k= \tan^{-1} 0.1$, $|k|=1$ m$^{-1}$,  the cell measurements $d=d_1=d_2=0.01$ m, and the aperture radius $a=0.001$ m.
In this case the parameter $\tau$ is complex and its magnitude is small, $\tau=5.0356 10^{-2}+i 5.1531 10^{-3}$.

The curves drawn in Figs 3 to 6 show the variation of the function $P(r,\Gt)$ with change of $\Gt$ when $r$ is kept constant.
In Fig. 3, we use  $\Gt_0=\pi/4$,   $|\Ga|=10$ m$^{-3}$, $\Gr_f/m_0=100$ m$^{-1}$,
  and $r=5$ m.
 For Fig. 4, we choose the same parameters   
 as for Fig. 3 except for $\Gt_0=\fr{\pi}{16}$. In Fig. 5, we increase the ratio $\Gr_f/m_0$
from 100 to 500 that results in a fivefold decrease of the panel surface density. The other parameters coincide with those 
employed for computations portrayed in Fig. 3. 
It is possible to infer from this figure that as the membrane surface density increases the absolute value
of the potential $\Gy_1$ is decreases. In Fig. 6, we decrease $r$ and select it to be $3$ m and keep the other parameters of Fig. 3
unchanged. 
Fig. 7 shows how the function $P(r)$ varies with change of $r$ when
the polar angle $\Gt$ equals $\pi/16$, $\pi/8$, $\pi/4$, $\pi/3$, and $5\pi/12$, while the 
other parameters are selected the same way as in Fig. 3. 
In Fig. 8, we increase the value
of $|k|$ from 1 to $5$, change its argument, $\arg k=\tan^{-1} 0.02$, and
because of formula (\ref{2.6}), increase $|\Ga|$ from 10 to 250.  The other parameters of Fig. 3 are the same.

As $k_2\to 0^+$ and $k_1\to k_{res}=\sqrt{2a/V}$, the parameter $\tau\to\infty$.
In Fig. 9, we change the cell measurements,  the aperture radius, and $\arg k$,
$d=d_1=d_2=0.2$ m, $a=0.005$ m and $\arg k=\tan^{-1}0.02$, and keep the other parameters the same as in Fig. 3. 
 In this case $\tau=0.96881+i0.17439$.
 It is seen that when $|\tau|$ is growing, the magnitude of the function $P(r)$ is also growing.
 We have  $P(r)\to\infty$ as 
 the wave number $k$ approaches the resonance value $k_{res}$.

\section{Conclusions} 

A closed-form solution has been given for the model problem of the scattering
of a plane sound wave by an infinite thin structure formed by a semi-infinite
acoustically hard screen attached to a sandwich panel with acoustically hard walls.
The upper side of the sandwich panel is perforated, while the lower side is an unperforated
membrane. We have applied two methods of extension of the boundary conditions to the whole 
real axis and deduce two order-2 vector Riemann-Hilbert problems. The matrix coefficients 
of both problems have
the Chebotarev-Khrapkov structure with the same order-4 characteristic polynomial but
with distinct entries. 
Wiener-Hopf matrix factors for both problems have been derived by quadratures by solving a
scalar Riemann-Hilbert problem on the same elliptic surface. The coefficient 
of the scalar  problem is equal to the first eigenvalue of the matrix on the upper sheet of the surface and
the second eigenvalue on the lower sheet. We have eliminated the essential singularity 
caused by simple poles of the Cauchy analogue at the two infinite points of the surface 
by solving a genus-1 Jacobi inversion problems in terms of the Riemann $\Gt$-function. 
 
We have found that the analysis of the Wiener-Hopf matrix factors at infinity is
simpler for the first method that sets the Riemann-Hilbert problem for the one-sided Fourier transforms
of the velocity potentials on the upper and lower sides of the infinite structure. The second method 
extends  the four boundary conditions to the whole real axis by means of unknown functions and 
employ the one-sided Fourier transforms of these functions.
The advantage of the first method over the second one is explained by the presence of the logarithmic growth at infinity 
of the densities of the singular integrals involved in the solution obtained by the second  
method. Both methods lead to the solution having two arbitrary constants. The constants have been 
fixed by additional conditions of the problem. 
For the first method, in addition to the meromorphic Wiener-Hopf factors, we constructed the canonical matrix
of factorization and computed the 
 partial indices of factorization. It turns out that they both are equal to zero and therefore stable.
  
 Numerical tests have been implemented for the solution derived by the first method.
 The integrals involved are rapidly convergent for all values of the parameters tested except for the case when 
 $|\tau|\to\infty$, when the method is not numerically efficient.
 We have computed the absolute values of the full velocity potentials, the function $P(r,\Gt)=|\Gf_1|$,
 $0<\Gt<\pi$, and $P(r,\Gt)=|\Gf_{inc}+\Gf_{ref}+\Gf_0|$, $\pi<\Gt<2\pi$. We have found that the presence 
 of the sandwich panel perforated from the upper side reduces the transmission of sound, and when the membrane surface density $m_0$ is growing
 the function $P(r,\Gt)$ ($0<\Gt<\pi)$ decreases. 
 We have also discovered that when the absolute value $|k|$ of the complex wave number approaches
 the resonance value $k_{res}$, then $|\tau|\to\infty$ and the magnitude of the function $P(r,\Gt)$ tends to infinity.

\vspace{.2in}

{\centerline{\Large\bf  References}}

\vspace{.1in}

\begin{enumerate}

\item\label{ffo}  J. E. Ffowcs Williams,  The acoustics of turbulence near sound absorbent liners,
\textit{J. Fluid Mech.}  \textbf{51} (1972) 737-749..

\item\label{lep0} F. G. Leppington and  H. Levine,
Reflexion and transmission at a plane screen with periodically arranged circular or elliptical apertures. 
\textit{J. Fluid Mech.}  \textbf{61} (1973) 109-127.

\item\label{lep} F. G. Leppington, The effective boundary conditions for a perforated elastic sandwich panel in
a compressible fluid,  \textit{Proc. R. Soc.}  A  \textbf{427} (1990) 385-399.

\item\label{jon} C. M. A. Jones, Scattering by a semi-infinite sandwich panel perforated on one side, \textit{Proc. R. Soc.}  A 
  \textbf{431} (1990) 465-479.

\item\label{as1}  Y. A. Antipov and V. V. Silvestrov, Factorization on a Riemann surface in scattering theory, \textit{Quart. J. Mech. Appl. Math.}   \textbf{55} (2002) 607-654.

\item\label{moi}  N.G.Moiseyev, Factorization of matrix functions of special form,  \textit{Soviet Math. Dokl.}   \textbf{39} (1989)
264-267.

\item\label{am}  Y. A. Antipov and N. G. Moiseyev, Exact solution of the plane problem for a composite plane
with a cut across the boundary between two media,   \textit{J. Appl.
Math. Mech. (PMM)}   \textbf{55} (1991) 531-539.

\item\label{che} G. N. Chebotarev, On closed-form solution of a Riemann boundary value problem for n pairs
of functions,  \textit{Uchen. Zap. Kazan. Univ.} \textbf{116} (1956) 31-58.

\item\label{khr} A. A. Khrapkov, Certain cases of the elastic equilibrium of an infinite wedge with a non- symmetric notch at the vertex, subjected to concentrated forces,  \textit{J. Appl. Math. Mech. (PMM)} \textbf{35} (1971) 625-637.

\item\label{dan}  V. G. Daniele, On the solution of two coupled Wiener–Hopf equations,  \textit{SIAM J. Appl. Math.}
  \textbf{44}  (1984) 667-680.

\item\label{abr} I. D. Abrahams, On the non-commutative factorization of Wiener–Hopf kernels of Khrapkov
type,  \textit{Proc. R. Soc. A} \textbf{454} (1998) 1719-1743.

\item\label{zve}  E. I. Zverovich, Boundary value problems in the theory of analytic functions in Ho\"lder classes on Riemann surfaces,  \textit{Russian Math. Surveys}  \textbf{26} (1971) 117-192.

\item\label{kra}  A. Krazer,  \textit{Lehrbuch der Thetafunktionen}, Teubner, Leipzig 1903.

\item\label{spr}  G. Springer,  \textit{Introduction to Riemann Surfaces}, Addison–Wesley, Reading, MA 1956.

\item\label{pap}  I.  Papanikolaou and F. G. Leppington,  Acoustic scattering by a parallel pair of semi-infinite wave-bearing surfaces,
 \textit{Proc. R. Soc.} 
A  \textbf{455} (1999) 3743-3765.

\item\label{dow}  A. P. Dowling and J. E. Ffowcs Williams,  \textit{Sound and Sources of Sound}, Ellis Horwood, Chichester, 1983.

\item\label{as2}  Y.A. Antipov and V.V. Silvestrov, Electromagnetic scattering from an anisotropic half-plane at oblique incidence: the exact solution, \textit{Quart. J. Mech. Appl. Math.}  \textbf{59} (2006)  211-251.

\item\label{gak} F. D. Gakhov, Riemann boundary-value problem for a system of n pairs of functions,  \textit{Russian
Math. Surveys}  \textbf{7} (1952) 3-54.

\item\label{mus} N. I. Muskhelishvili,  \textit{Singular Integral Equations}, Noordhoff, Groningen 1958.

\item\label{vek} N. P. Vekua,  \textit{Systems of Singular Integral Equations} Noordhoff, Groningen 1967.

\item\label{goh} I. Gohberg and M. G. Krein, On the stability of a system of partial indices of the Hilbert problem
for several unknown functions, \textit{Dokl. AN SSSR}  \textbf{119} (1958) 854-857.

 \end{enumerate}

 \end{document}